\documentclass[conference]{IEEEtran}
\IEEEoverridecommandlockouts
\usepackage{authblk}
\usepackage{cite}
\usepackage{amsmath,amssymb,amsfonts}
\usepackage{algorithmic}
\usepackage{graphicx}
\usepackage{textcomp}
\usepackage[table]{xcolor}
\usepackage{xcolor}
\usepackage{subcaption}
\usepackage{listings}
\usepackage{multicol}
\usepackage{tabularx}
\usepackage{multirow}
\usepackage{tikz}
\usepackage{eso-pic}

\usepackage{cleveref}
\usepackage{url}

\usepackage[outdir=./]{epstopdf}

\def\BibTeX{{\rm B\kern-.05em{\sc i\kern-.025em b}\kern-.08em
    T\kern-.1667em\lower.7ex\hbox{E}\kern-.125emX}}

\AddToShipoutPictureBG*{%
  \AtPageUpperLeft{%
    \put(0.07\paperwidth,-1.2cm){%
      \fbox{%
        \begin{minipage}{\textwidth}
          \centering
          \small
          \textbf{This paper has been accepted for publication in the proceedings of the IEEE/IFIP Network Operations and Management Symposium (NOMS) Conference, May 2024. The copyright for this paper will be transferred to IEEE.}
        \end{minipage}
      }%
    }%
  }%
}

\begin{document}

\title{\title{Quantifying the Impact of Frame Preemption on Combined TSN Shapers}}

\author[1]{Rubi Debnath}
\author[2]{Philipp Hortig}
\author[3]{Luxi Zhao}
\author[4]{Sebastian Steinhorst}

\affil[1,2,4]{TUM School of Computation, Information and Technology, Technical University of Munich, Germany}
\affil[3]{Beihang University, Beijing, China}
{
    \makeatletter
    \renewcommand\AB@affilsepx{, \protect\Affilfont}
    \makeatother

    \affil[1]{rubi.debnath@tum.de}
    \affil[2]{philipp.hortig@tum.de}
    \affil[3]{zhaoluxi@buaa.edu.cn}
    \affil[4]{sebastian.steinhorst@tum.de}
}

\maketitle

\begin{abstract}
Different scheduling mechanisms in Time Sensitive Networking (TSN) can be integrated together to design and support complex architectures with enhanced capabilities for mixed critical networks. Integrating Frame Preemption (FP) with Credit-Based Shaper (CBS) and Gate Control List (GCL) opens up different modes and configuration choices resulting in a complex evaluation of several possibilities and their impact on the Quality of Service (QoS). In this paper, we implement and quantify the integration of preemptive CBS with GCL by incorporating FP into the architecture. Our experiments show that the end-to-end delay of Audio Video Bridging (AVB) flows shaped by CBS reduces significantly (up to 40\%) when AVB flows are set to preemptable class. We further show that the jitter of Time Triggered (TT) traffic remains unaffected in "with Hold/Release" mode. Furthermore, we propose to introduce Guardband (GB) in the "without Hold/Release" to reduce the jitter of the TT flow. We compare all the different integration modes, starting with CBS with GCL, extending it further to FP. We evaluate all feasible combinations in both synthetic and realistic scenarios and offer recommendations for practical configuration methods.
\end{abstract}


\section{Introduction}
\label{sec:introduction}
In various domains, including smart factories, manufacturing, healthcare, autonomous vehicles, and industrial automation, real-time data communication is of paramount importance. Time-Sensitive Networking (TSN) comprises a range of sub-standards and mechanisms designed to provide real-time deterministic guarantees and diverse Quality of Service (QoS). Consequently, TSN has become a highly sought-after technology in these domains. The QoS of a TSN flow is measured by the end-to-end delay and the jitter of the flow while the delay of a TSN flow is predominantly influenced by several factors: interference from higher-priority traffic, interference from same-priority traffic, and interference from lower-priority traffic. To mitigate these interferences and ensure strict deadline guarantees, TSN has introduced specific mechanisms. One notable mechanism is the Time-Aware Shaper (TAS) \cite{8021Qbv}, which focuses on the temporal duration shared among gates. TAS operates based on parameters within the Gate Control List (GCL), which determine the open and close durations of Time-Triggered (TT) traffic gates. While TAS excels in providing strict deadline guarantees and zero jitter, the generation of the GCL poses a challenging problem known to be NP-hard.

Credit-Based Shaper (CBS) \cite{8021BA} is another shaping mechanism within TSN. CBS is specifically designed for Audio Video Bridging (AVB) flows to ensure bounded latency. TSN networks support the concurrent operation of multiple shapers on the same egress port. One widely discussed architecture in TSN is the coexistence of CBS with GCL, also known as CBS with TAS. In a CBS with GCL network, AVB flows can only be transmitted when the AVB queue gates are open, and there is no Guardband (GB). 

\begin{figure}[t!]
    \centering
        \includegraphics[scale=0.27,trim={0cm 9.5cm 1cm 0cm}, clip]{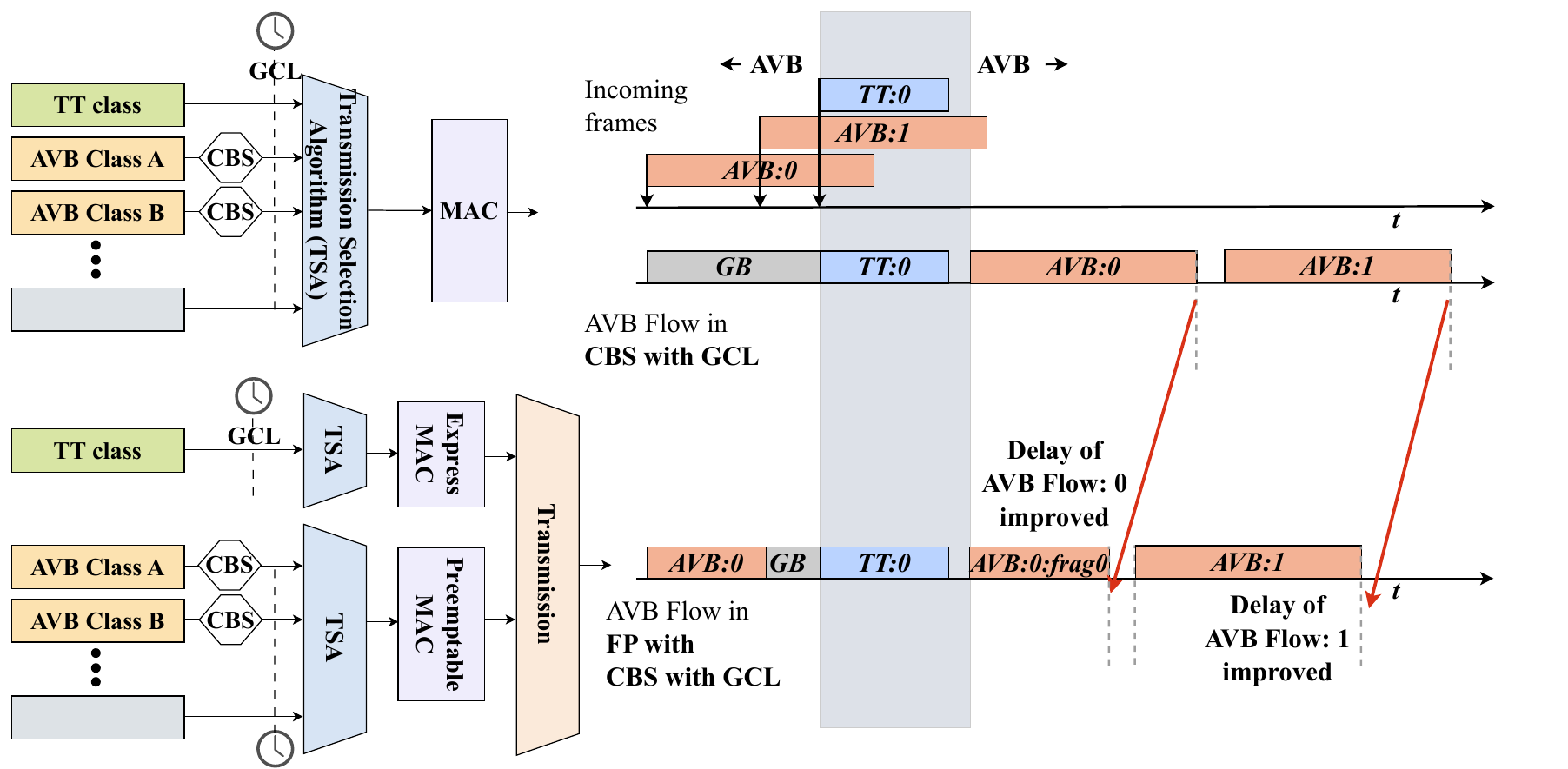}
        \caption{Motivational example of the benefits of using Frame Preemption in a complex TSN network.}
    \label{fig:motivation}
    \vspace{-0.6cm}
\end{figure}

In addition to TAS and CBS, another crucial mechanism in TSN is Frame Preemption (FP) \cite{8021Qbu} which enables higher-priority traffic to preempt lower-priority traffic transmissions, reducing interference in the network. In the FP mechanism, higher-priority traffic is categorized as "express traffic," allowing it to preempt "preemptable" traffic, which is generally made up of lower-priority traffic types. To comprehensively understand the advantages and disadvantages of using FP in a complex TSN network (as shown in Fig.~\ref{fig:motivation}), a thorough evaluation and analysis are essential. While related studies have assessed FP's performance, they often focus on isolated assessments, overlooking the broader impact of FP on various shapers and traffic types. Furthermore, the detailed exploration of using FP in conjunction with CBS and GCL is not well-analyzed in the literature. The primary objective of this paper is to thoroughly analyze and quantify the impact of FP when integrated with CBS and GCL.

\begin{figure}
    \centering
    \includegraphics[scale=0.22, trim={0cm 6cm 2cm 0cm}, clip]{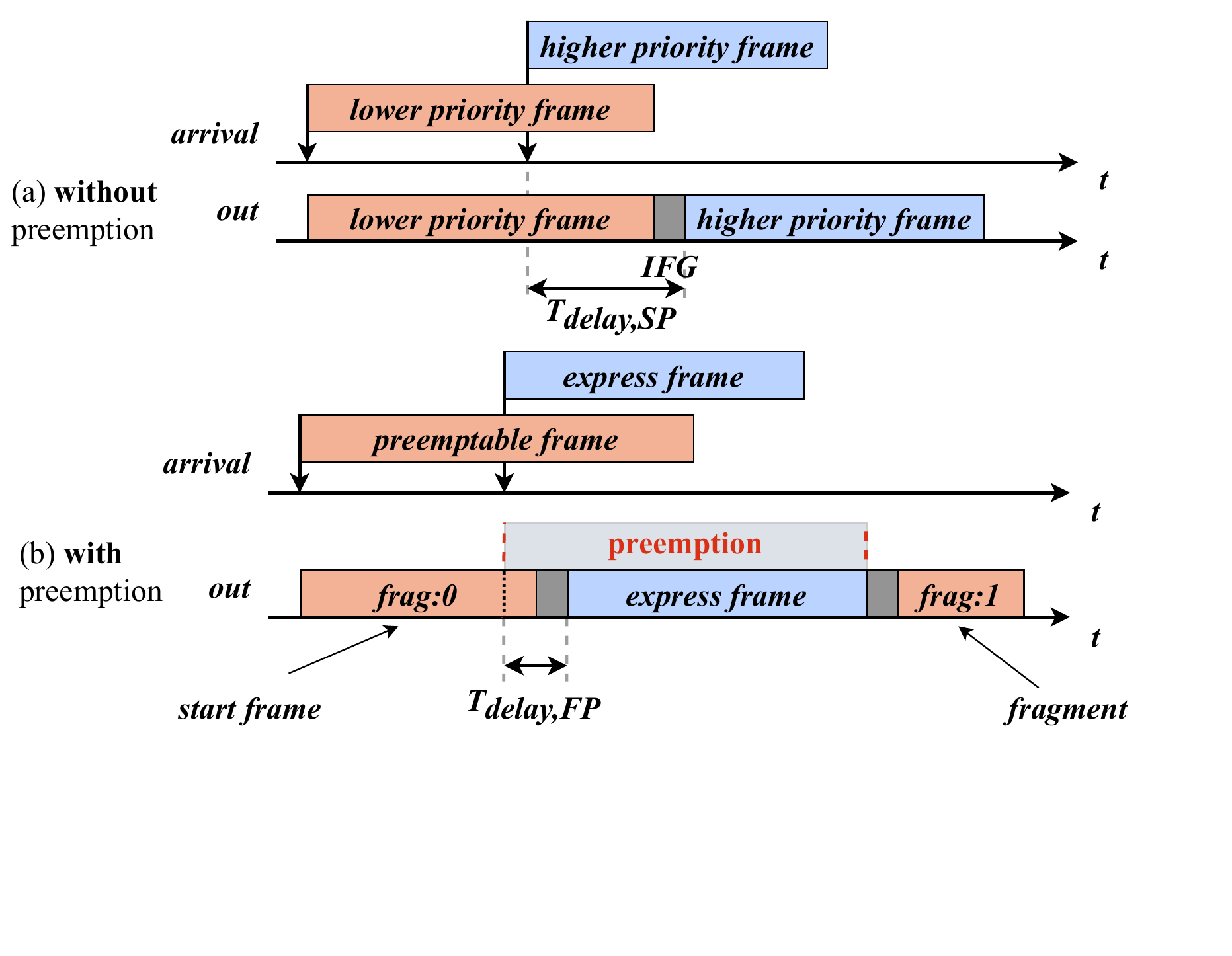}
    \caption{Basic Working Mechanism of (a) \textbf{no} FP (b) \textbf{with} FP.}
    \label{fig:preemption_basic}
    \vspace{-0.6cm}
\end{figure}

\subsection{Problem and Motivation}
\label{sub:motivation}
TSN provides diverse configuration options to accommodate different timing requirements. Existing literature provides individual insights on TAS, CBS, and FP. However, while the notion of enhancing AVB flow performance with FP in conjunction with CBS and GCL has been mentioned intuitively, it has not been comprehensively investigated. To bridge this research gap, we present a holistic approach encompassing the architecture, implementation, and evaluation of FP combined with CBS and GCL. Moreover, this paper explores the potential application of FP to enhance the performance of lower-priority non-Time-Triggered (non-TT) traffic within complex TSN networks. The key contributions of this paper are summarized as follows:

\textit{\textbf{Contribution 1:}} We implement and quantify the impact of FP on CBS and GCL and evaluate its performance in synthetic and realistic Orion Crew Exploration Vehicle (CEV)~\cite{luxi-cbs-multiple} test cases using OMNeT++\footnote{\url{https://omnetpp.org/}} and INET\footnote{\url{https://inet.omnetpp.org/2022-05-16-INET-4.4.0-released.html}}. To the best of our knowledge, this represents the first comprehensive analysis of the impact of FP on the CBS and GCL combination. Our computer simulations cover various CBS and GCL integration modes and incorporate different FP integration modes, providing the first complete support of FP with CBS and GCL. We have also made our implementation open source\footnote{\url{https://github.com/tum-esi/Quantifying-the-Impact-of-FP-on-CBS-GCL}} for further evaluation.

\textit{\textbf{Contribution 2:}} In this paper, we implement and show for the first time the enhanced performance of AVB in a CBS with GCL architecture when AVB flows are assigned to a preemptable class using FP (Section \ref{sec:evaluation}). Our results reveal a reduction of up to 40\% in Simulated Maximum Delay (SMD) of AVB traffic when placed in the preemptable class, thus significantly improving the performance of AVB flows. Importantly, we observed no detrimental effect on the jitter of the TT traffic in the FP \textbf{with Hold/Release} mode.

\textit{\textbf{Contribution 3:}} Our study outlines various integration modes of GCL and CBS in combination with FP, offering system engineers a wide array of options to choose from (Section \ref{sec:background} and \ref{sec:framework}). Choosing the right integration mode and its impact on the QoS is crucial (Section \ref{sec:evaluation}). Consequently, we present comprehensive results detailing the effects of different scheduling integration modes on both synthetic and realistic network topologies. Furthermore, the proposed architecture is versatile and can be adapted for use with other TSN traffic types or TSN shapers.

\section{Background}
\label{sec:background}
TAS ensures the meeting of timing requirements by providing temporal isolation for different traffic types, offering deterministic guarantees to meet flow deadlines. In contrast, CBS operates on the principle of "credit" and during the transmission of an AVB flow, the "credit" decreases with the $sendSlope$. The $sendSlope$ is calculated as given below:

\vspace{-0.59cm}
\begin{align}
  sendSlope = idleSlope - portTransmitRate
\end{align}
\vspace{-0.49cm}

where $portTransmitRate$ denotes the transmission rate in bits per second (bps) of the respective egress port, while $idleSlope(N)$ specifies the slope of $credit$ increase immediately after the frame transmission ends. When the associated AVB queue is empty, the $credit$ is set to zero. In this paper, we assume that the $idleSlope(N)$ (the value of the reserved Bandwidth (BW)) is predefined by the network designer. When there are multiple AVB classes, ranging from $1$ to $N$, each AVB class, denoted as $N$, is assigned an individual $idleSlope(N)$. Furthermore, within our framework, the total reserved bandwidth for AVB flows is consistently set at 85\% of the available $portTransmitRate$, taking into consideration the available AVB load in the network.
In a TSN network with both CBS and GCL, TT and AVB traffic gates are mutually exclusive. When the TT gate is open, the AVB gate is closed, and vice versa. This means AVB frames can't be sent when the AVB gate is closed, and vice versa. When AVB flows are in the queue, waiting during TT flow transmission, the credit behavior of CBS has various possible implementations, also referred to as the integration modes of CBS and GCL. 

The credit integration mode is crucial as it significantly impacts the credit recovery of the CBS algorithm and, consequently, the performance of the AVB flows \cite{boyer-cbs-frozen}. The discussion of the "credit" evolution focuses on two situations: 1) the gate closing period and 2) the active GB before the gate closing period. During the closure of the AVB gates, credit accumulation is paused. However, the credit behavior during the active GB before the gate closing period has various possible implementations. The IEEE 802.1Q-2018 standard~\cite{8021BA} recommends pausing credit accumulation during the closure of the associated AVB gate. While the term 'frozen' is not explicitly used in the standard, it is employed in related works like \cite{boyer-cbs-frozen, luxi-cbs-GB, luxi-cbs-multiple} to describe the credit behavior during the GB. In this paper, the terms 'frozen' and 'nonfrozen' credit are specifically used to describe the credit behavior during the GB. The behavior of pausing credit during AVB gate closure is consistent across all the different integration modes.

The basic working principle of the FP mechanism is illustrated in Fig. \ref{fig:preemption_basic}\textbf{(b)}. FP works by dividing traffic types into two different groups: \textit{express traffic} and \textit{preemptable traffic}. \textit{Express Traffic} refers to traffic with higher priority that cannot be interrupted, while \textit{preemptable traffic} includes traffic types that can be preempted and interrupted. The assignment of traffic types to the \textit{express} and \textit{preemptable} classes is a design choice made by the system engineer. When an \textit{express} frame is ready for transmission and a \textit{preemptable} frame is in the transmission process, the \textit{express} traffic preempts the \textit{preemptable} traffic. When preemption is triggered by an \textit{express} frame, the \textit{preempted} frame must be completed with a $4Byte$ Frame Check Sequence (FCS) and $12 Byte$ for the Interframe Gap (IFG). Considering a link rate of $100 Mbps$, this results in a processing overhead of $T_{delay,FP} = 1.28 \mu s$ for preemption delay.

\section{Proposed Framework}
\label{sec:framework}
In our model, we have both TT and AVB flows. TT traffic is scheduled using the GCL, while AVB flows are shaped using CBS and are transmitted only when the AVB gates are open. AVB flows are categorized as preemptable and can be preempted using FP.

\subsection{CBS with GCL}
In a CBS with GCL architecture, many AVB frame transmissions become impossible due to a closed AVB queue gate or an active GB. This cumulative unavailability of the link to AVB flows necessitates the application of the following equation for calculating the $idleSlope$.

\vspace{-0.55cm}
\begin{align}
\label{eq:operidleslope}
    idleslope(N) &= operIdleSlope(N) \cdot \frac{OperCycleTime}{GateOpenTime(N)}
\end{align}

The $OperCycleTime$ and $GateOpenTime(N)$ is further expressed as follows. 

\vspace{-0.5cm}
\begin{align}
\label{eq:operidleslope}
    idleslope(N) &= operIdleslope(N) \cdot \\ 
    & \left(\frac{hyperperiod_{GCL}}{hyperperiod_{GCL} - T_{GateClosed}}\right)
    \nonumber \nonumber 
\end{align}

where $operIdleSlope(N)$ represents the original BW value for AVB class $N$, $hyperperiod_{GCL}$ corresponds to the cycle duration of the GCL, and $T_{GateClosed}$ signifies the duration for which the transmission gate for AVB class $N$ remains closed during one cycle. To obtain a more precise value, it is essential to account for the time of wasted BW, which arises due to the GB and is denoted as $T_{GB}$.

\vspace{-0.5cm}
\begin{align}
    idleslope(N) &= operIdleslope(N) \cdot \\ 
    & \left(\frac{hyperperiod_{GCL}}{hyperperiod_{GCL} - T_{GateClosed} - T_{GB}}\right)
\nonumber \nonumber 
\label{eq:second_equation}
\end{align}

\begin{figure}[t]
    \centering
    \includegraphics[scale=0.24, trim={0cm 4.5cm 2cm 0cm}, clip]{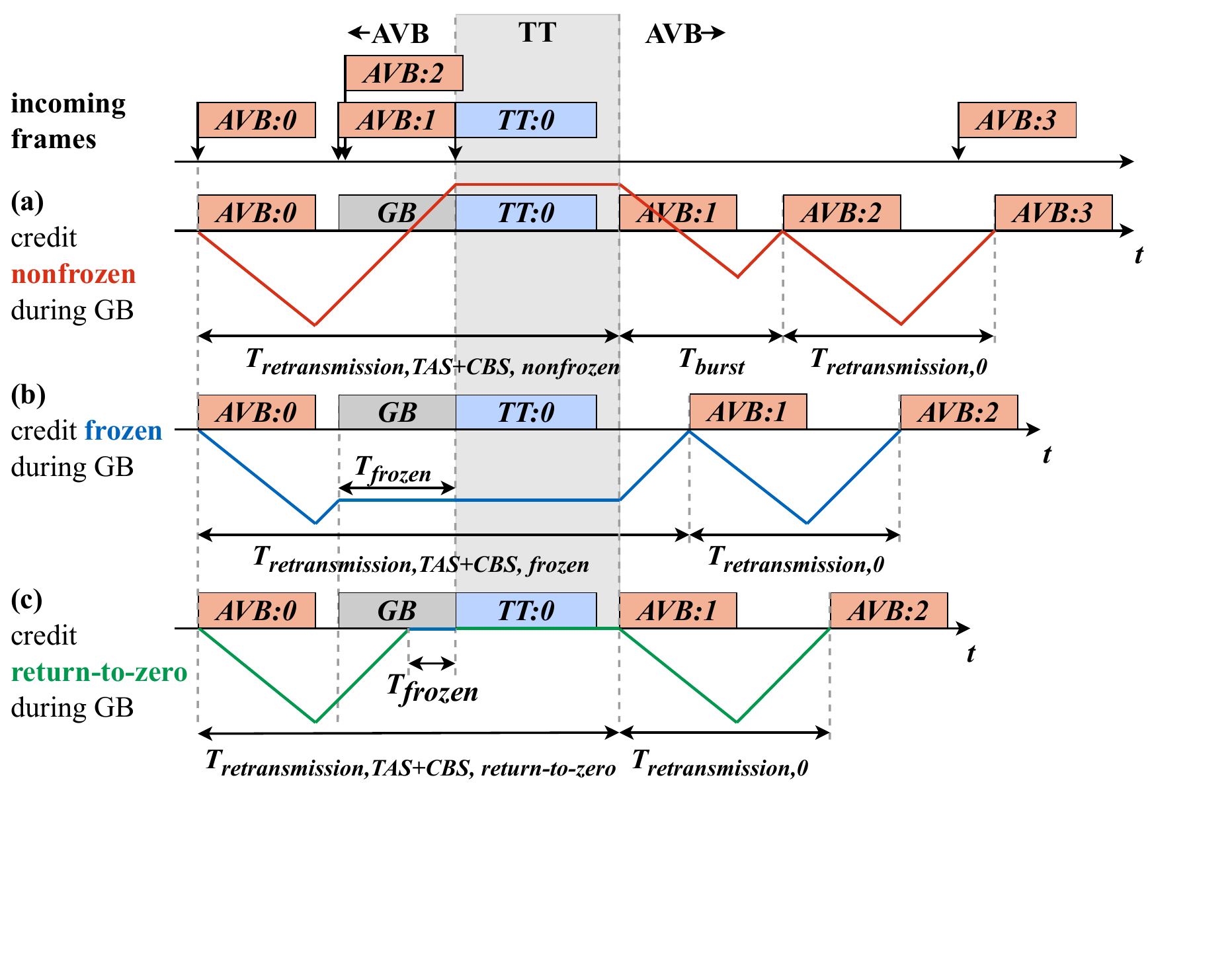}
    \caption{\textit{CBS with GCL} Credit Integration Modes: Credit evolution (a) \textbf{nonfrozen} (b) \textbf{frozen} and (c) \textbf{return-to-zero} during GB.}
    \label{fig:TAS_CBS_credit_integration_modes}
    \vspace{-0.5cm}
\end{figure}

\begin{figure}[t]
    \centering
    \includegraphics[scale=0.25, trim={0cm 1cm 0cm 1.5cm}, clip]{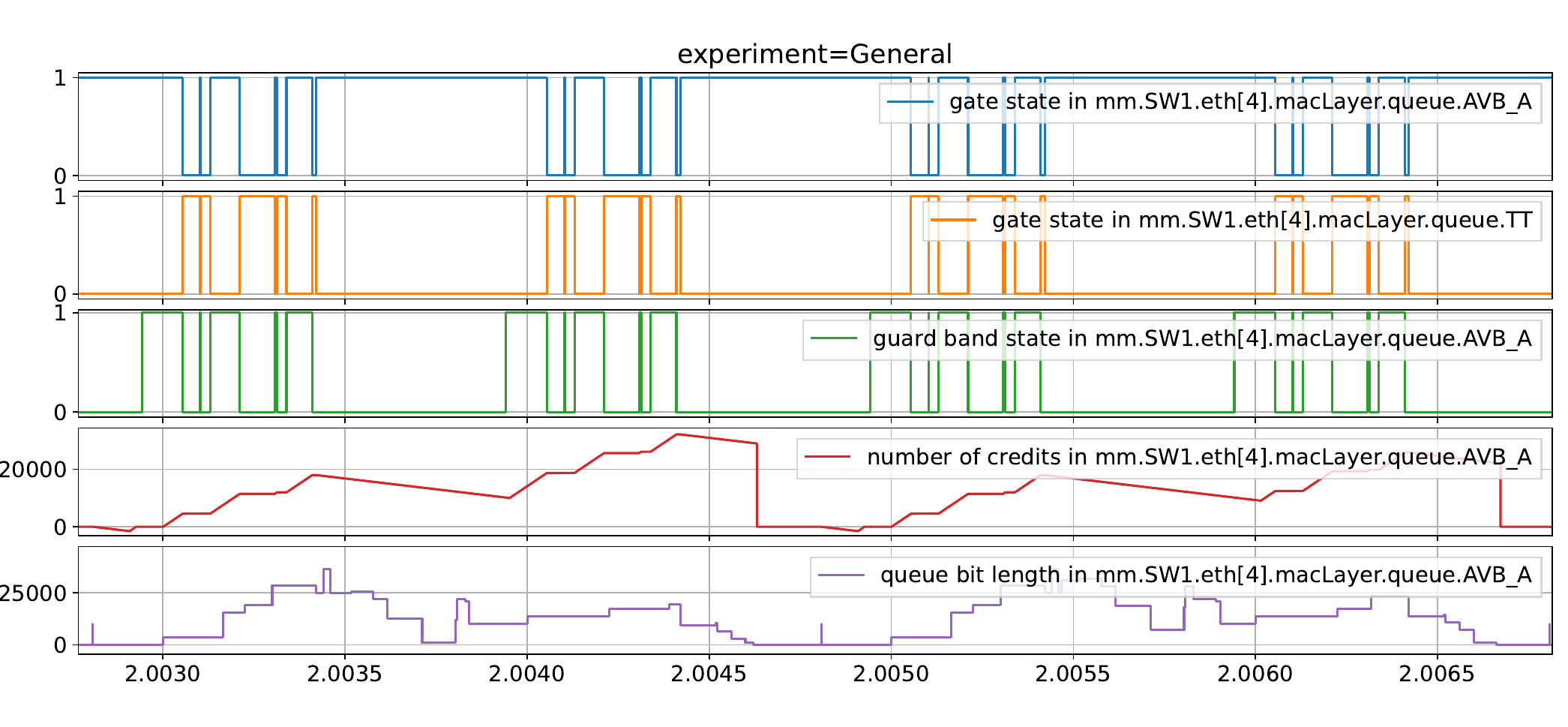}
    \caption{Simulation of the \textbf{Nonfrozen} mode showcasing the credit behavior while the GB is active.}
    \label{fig:standard_credit_behaviour}
    \vspace{-0.5cm}
\end{figure}

\begin{figure}[t]
    \centering
    \includegraphics[scale=0.25, trim={0cm 1cm 0cm 1.5cm}, clip]{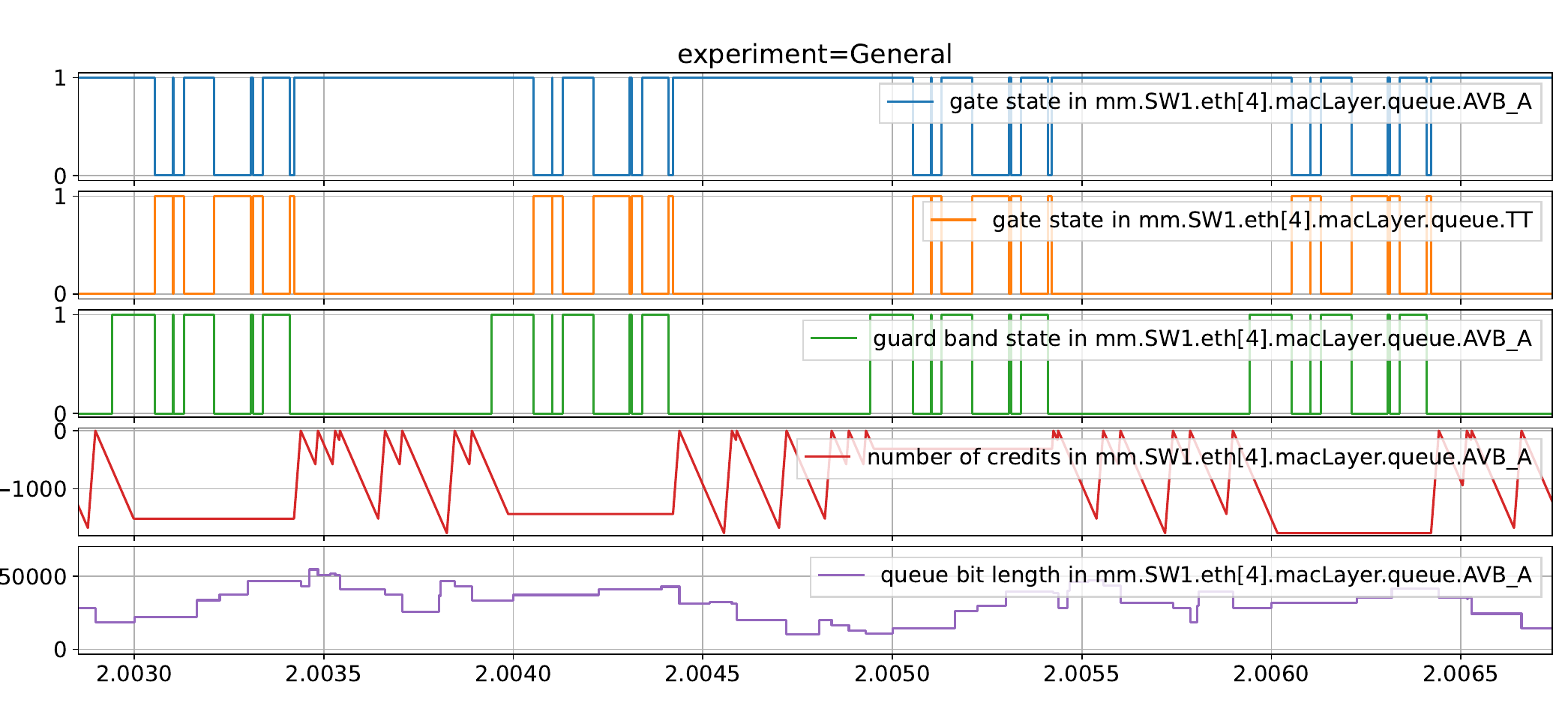}
    \caption{Simulation of the \textbf{Frozen} mode showcasing the credit behavior.} 
    \label{fig:frozen_credit_behaviour}
    \vspace{-0.5cm}
\end{figure}

\begin{figure}[t]
    \centering
    \includegraphics[scale=0.28, trim={0cm 1cm 0cm 1.3cm}, clip]{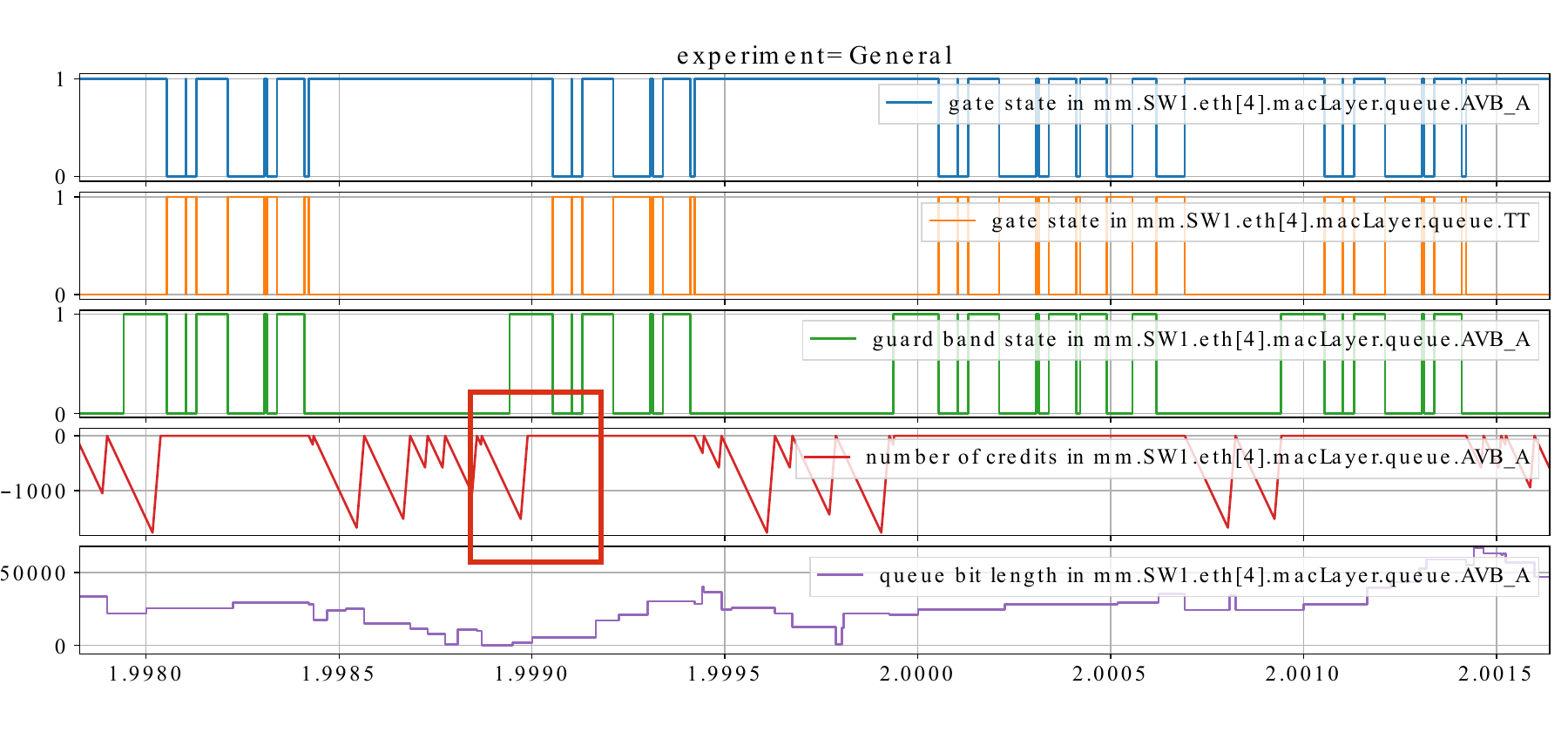}
    \caption{Simulation of the \textbf{Return-to-Zero} mode showcasing the credit behavior.}
    \label{fig:returnToZero_credit_behaviour}
    \vspace{-0.3cm}
\end{figure}

The various "credit" modes within the CBS with GCL architecture are further explained, and the simulation of "credit" evolution is discussed.

\subsubsection{Nonfrozen Credit During GB}
\label{sec:standard_credit_behaviour}
During the GB, AVB transmission is prohibited to avoid the tail from a previous AVB frame protruding into the subsequent time window of TT frame. In the \textit{nonfrozen} mode, illustrated in the block diagram in Fig. \ref{fig:TAS_CBS_credit_integration_modes} (\textbf{a}), the credit continues to rise during the active GB. A key distinction in this approach is the accumulation of positive "extra bonus credit" during the GB period. After the AVB gate reopens, this positive "credit" accumulation results in shorter retransmission duration for AVB frames, leading to quicker queue backlog reduction and, consequently, smaller Worst-Case Delay (WCD) for AVB flows. Furthermore, AVB traffic classes with high accumulated bonus credit take precedence over other coexisting lower-priority traffic classes during this time window, as previously noted by Boyer et al. in \cite{boyer-cbs-frozen}.

\subsubsection{Frozen Credit During GB}
\label{sec:frozen_credit_during_guardband_behavior}
As depicted in Fig. \ref{fig:TAS_CBS_credit_integration_modes} (\textbf{b}), in the \textit{frozen} mode, the credit freezes during an active GB. The "credit" value in this diagram is initially negative and only starts increasing with the \textit{idleSlope} after the associated AVB gate reopens. However, as shown in Fig. \ref{fig:TAS_CBS_credit_integration_modes}(\textbf{b}), in this example scenario, AVB frame transmission cannot occur immediately after the AVB gate opens due to insufficient available credit. Credit accumulates only after the gate opens, extending the credit recovery time. In contrast to the "nonfrozen" mode, the "frozen" mode doesn't result in a substantial positive credit value. This, however, comes at the cost of a longer waiting time for AVB frames in the queue, leading to a higher WCD. 

\subsubsection{Return-To-Zero Credit During GB}
\label{sec:return_to_zero_credit_during_guardband}
Boyer et al. proposed in \cite{boyer-cbs-frozen} the \textit{return-to-zero} credit behavior. In \textit{return-to-zero}, the credit returns to the zero value during the \textit{GB}. On one hand, this provides the advantage of having instant transmission after gate reopening, and on the other hand, it further prevents the credit from increasing to high positive values during \textit{GB}. As none of the related work evaluated this mechanism, in this paper, we further implemented and evaluated this mechanism to cover all the credit modes.

\subsubsection{\textbf{Quantitative Analysis of different "credit" modes}}
We have quantitatively demonstrated the different "credit" evolution behaviors through our simulation model. In Fig. \ref{fig:standard_credit_behaviour}, \ref{fig:frozen_credit_behaviour}, and \ref{fig:returnToZero_credit_behaviour}, the red curve represents the "credit" in Mbps, the TT gate state is indicated in blue, the AVB Class A gate state is shown in orange, the GB state is depicted in green, and the queue backlog is displayed in violet. Fig. \ref{fig:frozen_credit_behaviour} highlights a significantly higher AVB queue backlog in the "frozen" mode, contributing to increased WCDs of AVB flows in the "frozen" integration mode, as discussed in Section \ref{sec:evaluation}. This effect is more pronounced in multi-hop networks, where each node adds additional delay. Conversely, in the "nonfrozen" mode, as shown in Fig. \ref{fig:standard_credit_behaviour}, the AVB queue backlog is substantially lower.

\begin{figure}[t!]
    \centering
    \includegraphics[scale=0.26, trim={1cm 11.7cm 2cm 0cm}, clip]{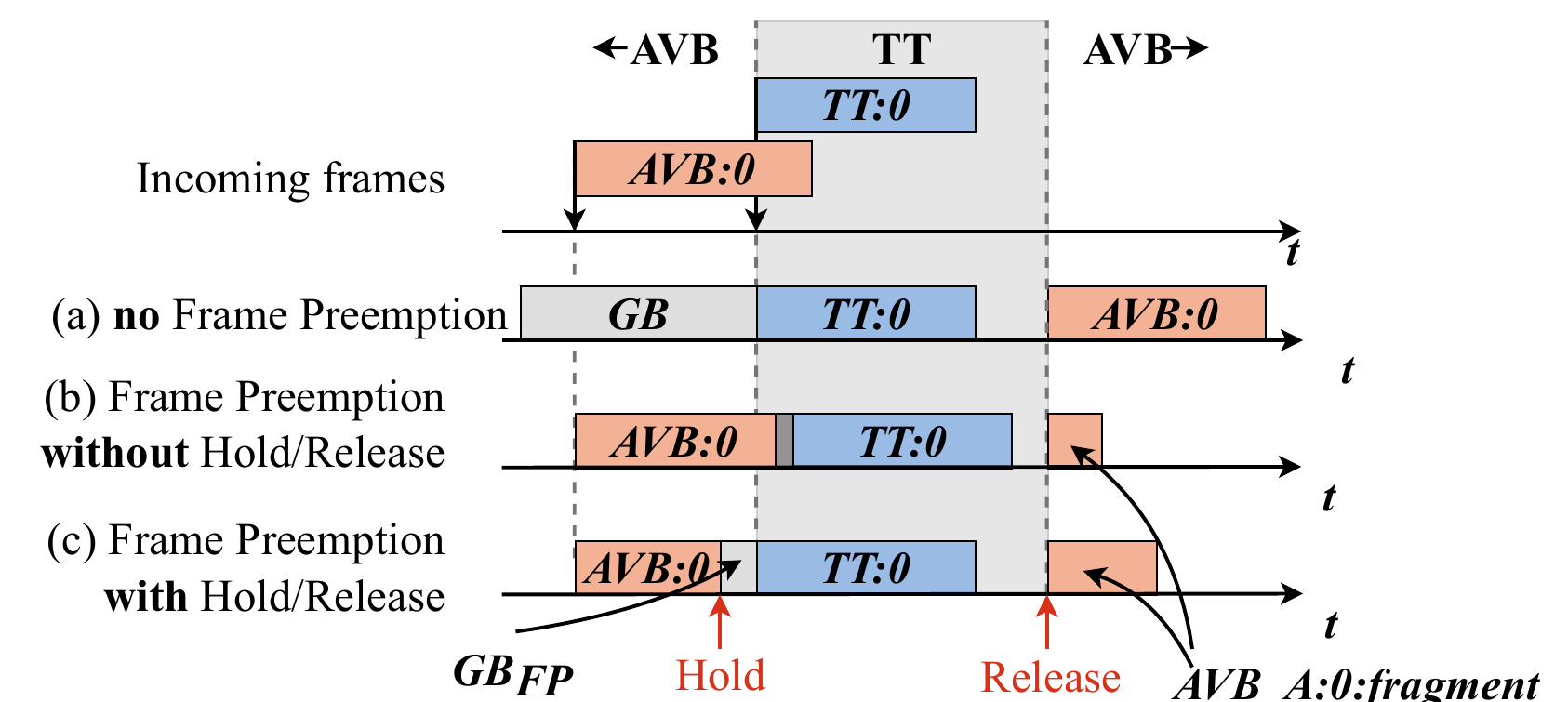}
    \caption{TAS (a) without preemption (b) with FP without Hold/Release mode and (c) with FP with Hold/Release mode.}
    \label{fig:frame_sequence_chart_TAS_FP}
    \vspace{-0.6cm}
\end{figure}

\subsection{FP with CBS with GCL}
To protect TT traffic from other interfering traffics, a GB is used. One reason for the high delay of AVB is the GB, during which the AVB frames cannot be transmitted. Reducing the size of the GB would result in a reduction of the WCD of the AVB flows and higher BW utilization \cite{voica}. This paper proposes the use of FP to preempt AVB flows, which are shaped using CBS, while also reducing the GB size. This approach enhances AVB performance, especially since FP introduces minimal overhead. We explore two FP integration modes: FP with Hold/Release and FP without Hold/Release, illustrated in Fig.~\ref{fig:frame_sequence_chart_TAS_FP}. When we combine FP with CBS and GCL, we also combine FP with Hold/Release and FP without Hold/Release with the frozen, nonfrozen, and return-to-zero modes. With the inclusion of FP, it becomes necessary to update the credit integration modes in the said combination context. To the best of our knowledge, such a detailed combination of all integration modes has not been implemented or evaluated previously.

\begin{figure}[t!]
    \centering
    \includegraphics[scale=0.24, trim={0cm 6cm 0cm 0cm}, clip]{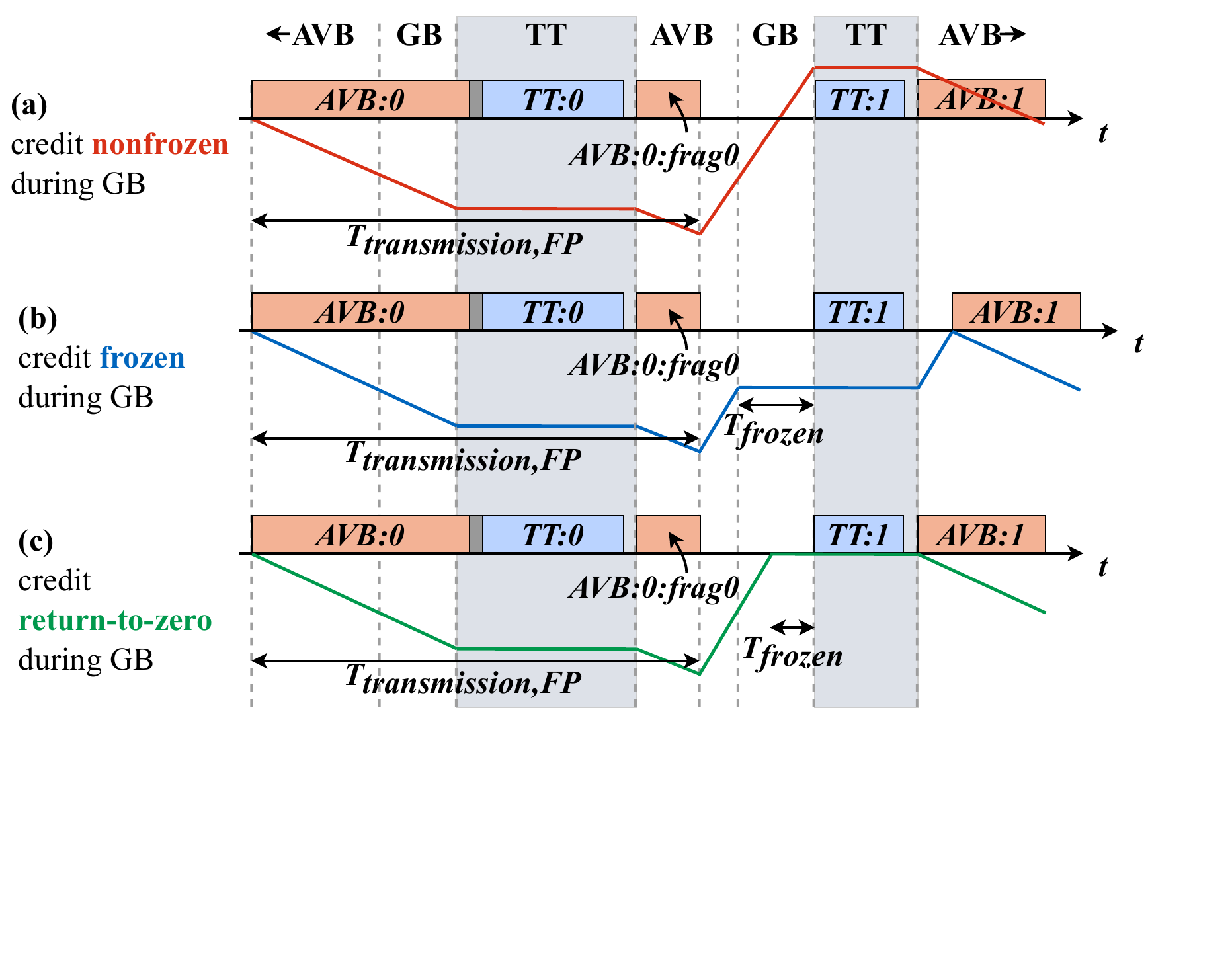}
    \caption{Credit evolution of (a) nonfrozen (b) frozen (c) return-to-zero for CBS with GCL \textbf{without} Hold/Release.}
    \label{fig:TAS_CBS_FP_withoutHR}
    \vspace{-0.69cm}
\end{figure}

\begin{figure}[t!]
    \centering
        \includegraphics[scale=0.24,trim={0cm 2cm 1cm 0cm}, clip]{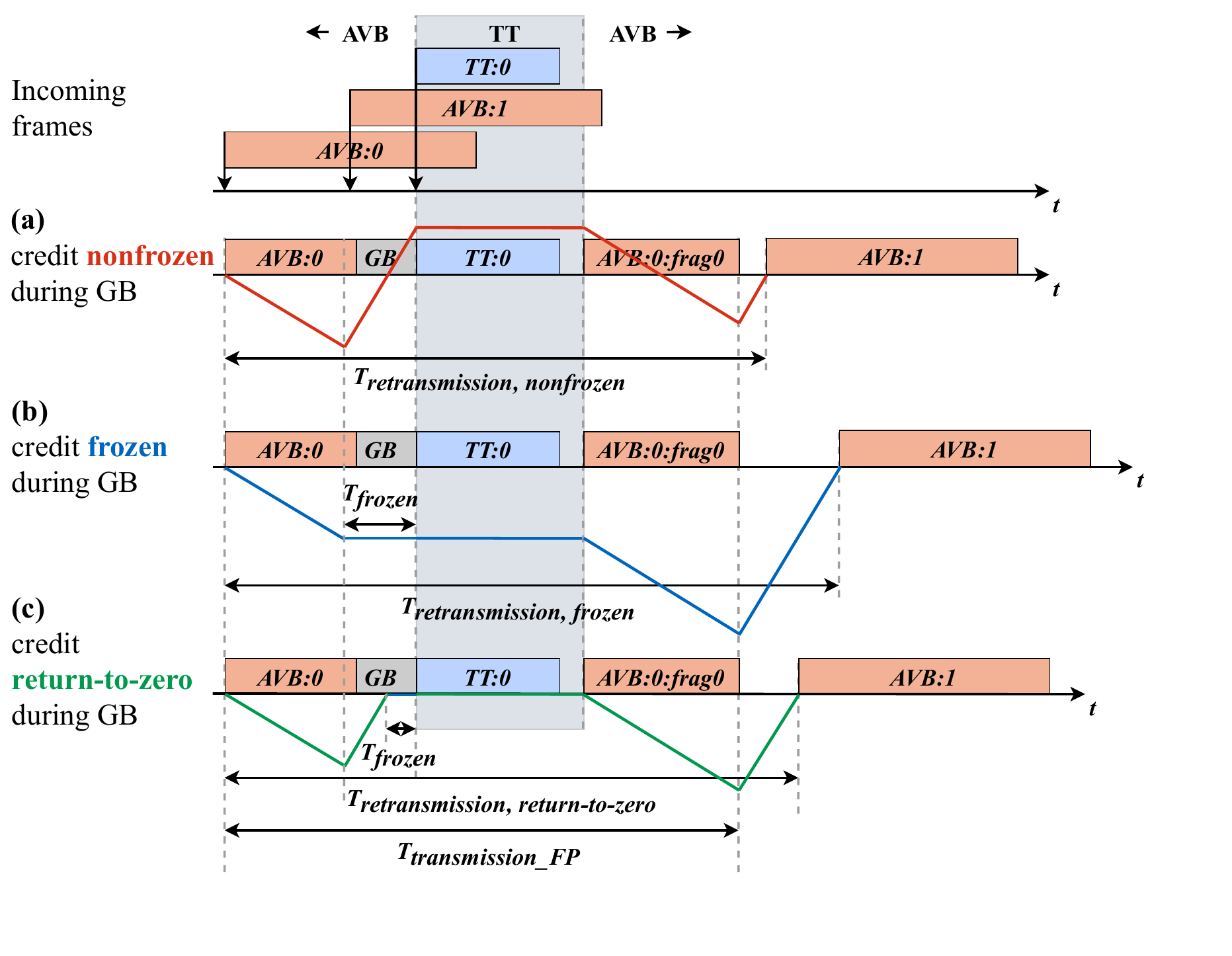}
        \caption{Credit evolution of (a) nonfrozen (b) frozen (c) return-to-zero integration modes for FP with CBS with GCL \textbf{with} Hold/Release.}
    \label{fig:TAS_CBS_FP_credit_integration_modes}
    \vspace{-0.6cm}
\end{figure}

\subsubsection{FP without Hold/Release}
\label{sec:preemption_without_hold_release}
In the "without Hold/Release" mode, the preemptable traffic gate close event triggers the preemption of the preemptable traffic. After the initiation of the preemption, preemptable traffic takes a certain amount of time to finish the transmission of the current frame that is under transmission before the express frame starts transmitting. Fig.~\ref{fig:TAS_CBS_FP_withoutHR} shows the credit evolution of CBS for "nonfrozen", "frozen", and "return-to-zero" for FP without Hold/Release mode.\\ \textbf{\textit{Modifications in "without Hold/Release" proposed in this paper:}} IEEE 802.1Q \cite{8021Q} does not mention the GB in the "without Hold/Release" mode. However, the absence of a GB can lead to delays in TT traffic, resulting in high jitter. For instance, if there is no GB, and an AVB frame with a 123 Byte payload dequeues shortly before the TT time window opening, but the TT frame to be transmitted is only 80 Bytes, this can create a problem. In this scenario, the AVB frame cannot be preempted because its size is 123 Bytes. Consequently, the entire AVB frame is transmitted, causing the TT traffic to miss its transmission window. To address this issue, we propose and implement the GB in the "without Hold/Release" mode in this paper.

\subsubsection{FP with Hold/Release}
\label{sec:preemption_with_hold_release}
To avoid delays caused by the preemption overhead on express traffic, FP with Hold/Release initiates the preemption mechanism early enough so that the preemption process of the preemptable traffic is completed before the express traffic gate opens. The minimum frame length assigned to the preemptable class determines how much the preemption is advanced in time. Fig.~\ref{fig:TAS_CBS_FP_credit_integration_modes} illustrates the credit evolution of CBS for the nonfrozen, frozen, and return-to-zero modes in FP with Hold/Release.

\begin{figure}[t]
    \centering
    \includegraphics[scale=0.15, trim={1cm 7cm 2cm 0cm}, clip]{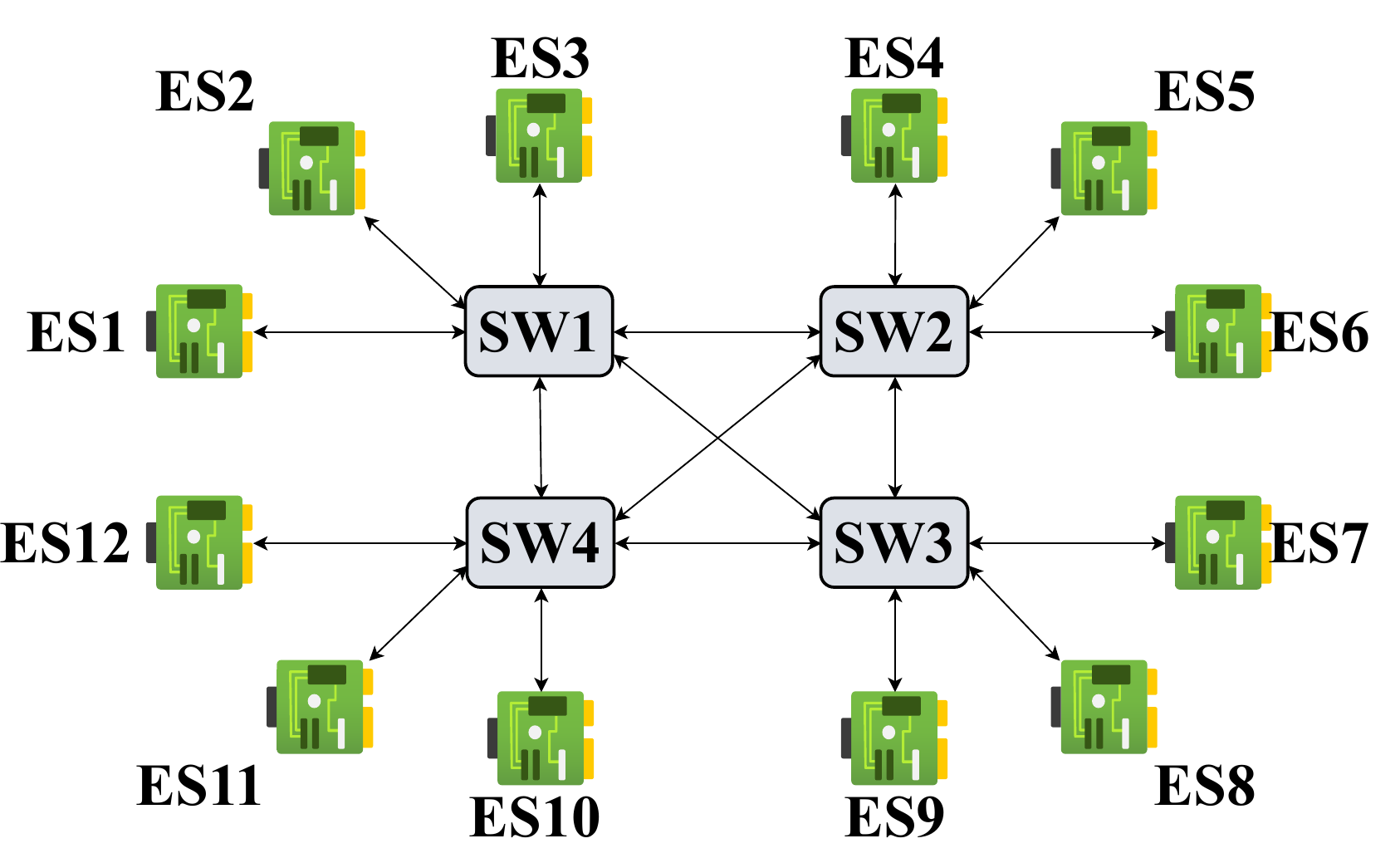}
    \caption{Medium Mesh Topology.}
    \label{fig:medium_mesh_topology}
    \vspace{-0.46cm}
\end{figure}

\begin{figure}[t]{}
\centering
\includegraphics[scale=0.24, trim={0cm 6.9cm 2cm 0cm}, clip]{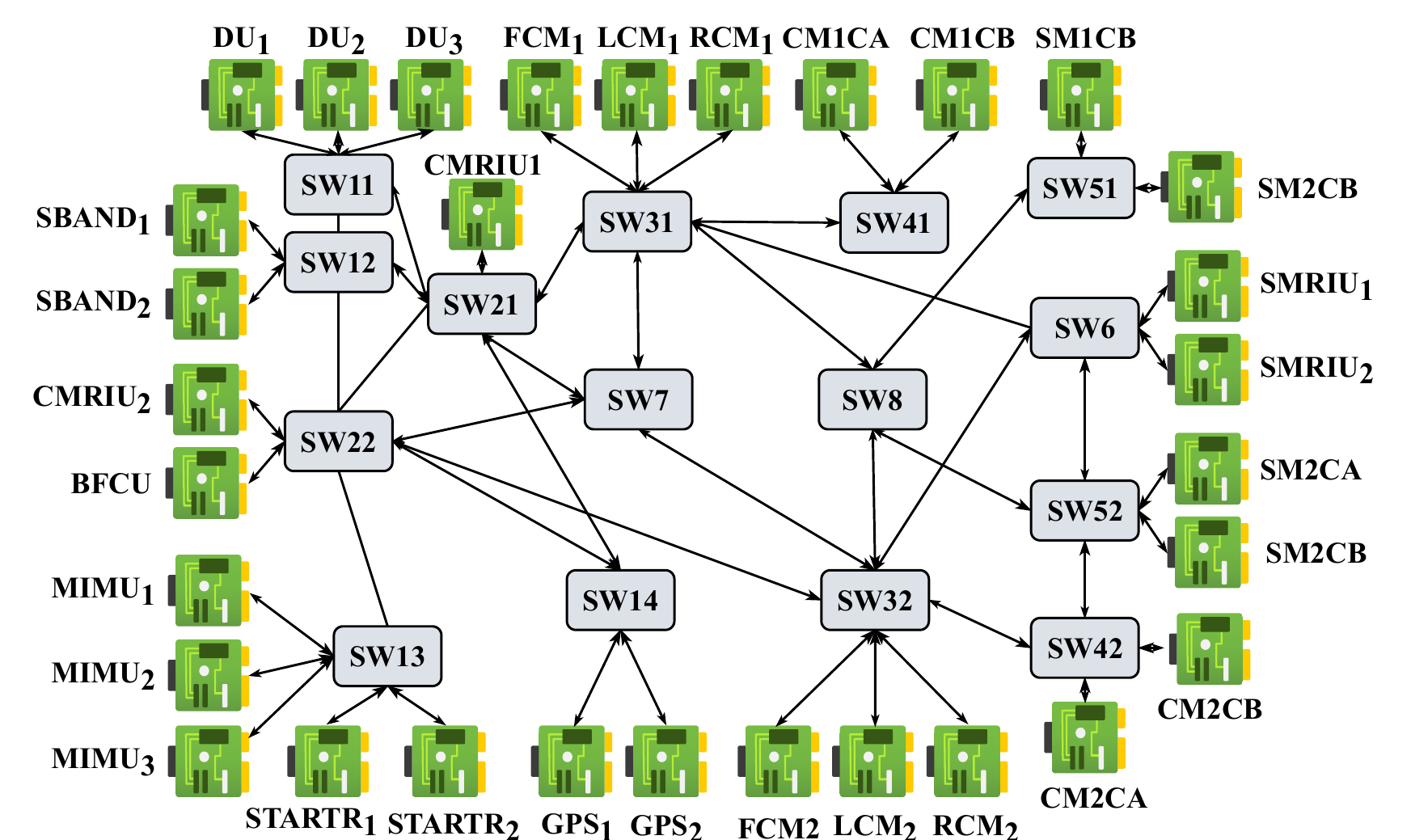}
\caption{Orion topology.}
\label{fig:orion_topology}
\vspace{-0.58cm}
\end{figure}

\section{Implementation and Simulation Model}
\label{sec:model}
In this paper, the TSN network consists of End Stations ($ESs$) and Switches ($SWs$). Both ESs and SWs are referred to as nodes, connected via full-duplex Ethernet links. The Ethernet link speed is configurable and tested under two widely used link capacities, 100 Mbps and 1 Gbps. The TSN network is modeled as an undirected graph $G(V,E)$, where $V = ES \cup SW$, with $V$ being a set of the nodes ($ESs$ and $SWs$) in the network, and $E$ being the set of edges (or connected links). $SWs$ support the GCL, CBS, and FP functionalities. All traffic types are configurable and can be set to either \textit{express} or \textit{preemptable}.

\vspace{-0.2cm}
\subsection{Implementation}
We used OMNeT++ and INET 4.4 library for the implementation in this paper. Although basic CBS and FP functionality exists in INET 4.4, the complex integration of "CBS with GCL" and "FP with CBS with GCL" is not supported. Therefore, we first implemented the \textit{frozen}, \textit{nonfrozen}, and \textit{return-to-zero} modes to support CBS with GCL. After that, to support FP in combination with CBS and GCL, we implemented FP "with" and "without" "Hold/Release" mechanisms to integrate FP with other shapers. In our model, the preemption mechanism no longer solely depends on the availability of an express frame but also takes into account the transmission gate state (open or closed) and the GB state. Here, FP is triggered by changes in the transmission gate state (for "without Hold/Release") or the GB state (for "with Hold/Release") of the express traffic. 
Furthermore, in our implementation, GCL is utilized for TT traffic (mapped to the express class), while CBS serves as the shaping mechanism for AVB flows, which are assigned to the preemptable class. Our model offers configurability for all these different modes, enabling us to designate any traffic type as preemptable or express class.

\begin{figure}[t!]
    \centering
    \includegraphics[scale=0.25, trim={0.5cm 0.5cm 3cm 1.2cm}, clip]{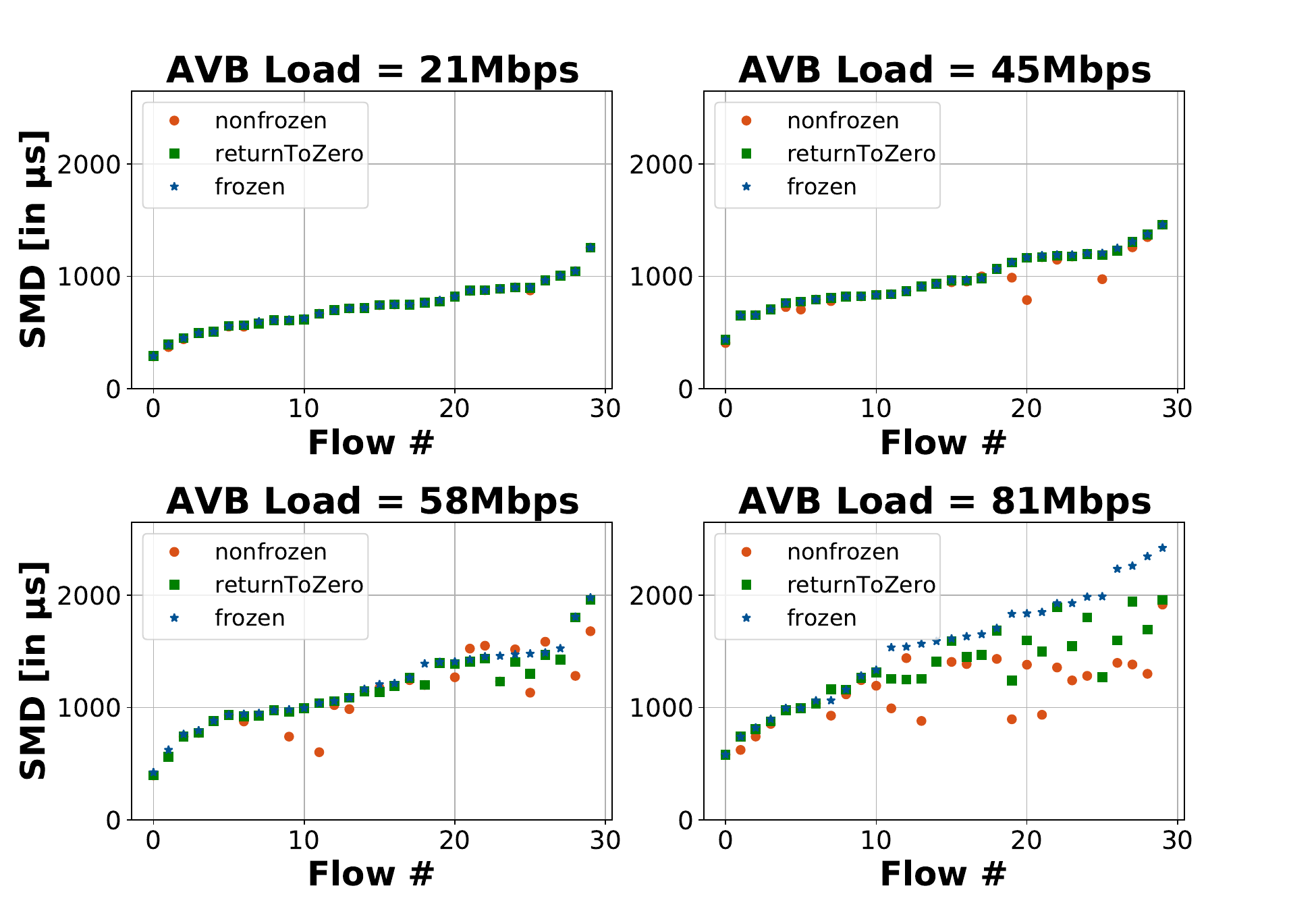}
    \caption{\textbf{MM} Topology: \textbf{SMD} of AVB for \textbf{CBS with GCL.}}
    \label{fig:SMD_AVBA_30to60_frozen_nonfrozen}
    \vspace{-0.5cm}
\end{figure}

\vspace{-0.2cm}
\subsection{Simulation Parameters}
The hardware and processing delays of the $SWs$ are set to 4$\mu$s, and Ethernet cable lengths between the nodes are 10m. Simulations were conducted on an Intel Core i7 4-core CPU with 16GB of RAM and the Windows 11 Operating System. AVB flow start times were randomly selected to introduce interference, and each test case was run 10 times for various network loads. TSN flow payload sizes ranged randomly from 64 to 1500 Bytes. Simulations in OMNeT++ had a duration of 30 seconds on the simulation clock.

\vspace{-0.1cm}

\subsection{Key Performance Indicators (KPIs)}
\label{sec:kpis}

\subsubsection{Simulated Maximum Delay (SMD)}
It is the time required for a frame to travel from the sending node to a receiving node in the network. In this paper, the SMD is calculated from the moment a frame is sent by the sending node until it is fully received by the destination node.

\subsubsection{Simulated Maximum Jitter (SMJ)}
SMJ denotes the maximum latency variation for each flow during the simulation. In this paper, we calculate SMJ for each flow by first computing the differences between consecutive end-to-end delay values and then extracting the maximum absolute difference.

\subsection{Topology and Test Cases}
\label{sec:topology}
We used the Medium Mesh (MM) topology shown in Fig.~\ref{fig:medium_mesh_topology} to test the synthetic test case. For the real-world use case, we used the ORION CEV Topology shown in Fig.~\ref{fig:orion_topology}, which is widely used in TSN evaluation. The test cases of the Orion Topology are taken from \cite{luxi-cbs-multiple}, and the test cases of the MM topology are taken from \cite{luxi}. Both network topologies consist of TT and AVB flows. The link capacity of the Orion topology is set to 1 Gbps, and for MM, it is set to 100 Mbps.

\section{Experimental Evaluation}
\label{sec:evaluation}
We first evaluate the integration modes of CBS and GCL, and then assess the impact of FP on CBS and GCL.

\begin{figure}[t!]
    \centering
    \includegraphics[scale=0.25, trim={0.5cm 0.4cm 0.7cm 1.5cm}, clip]{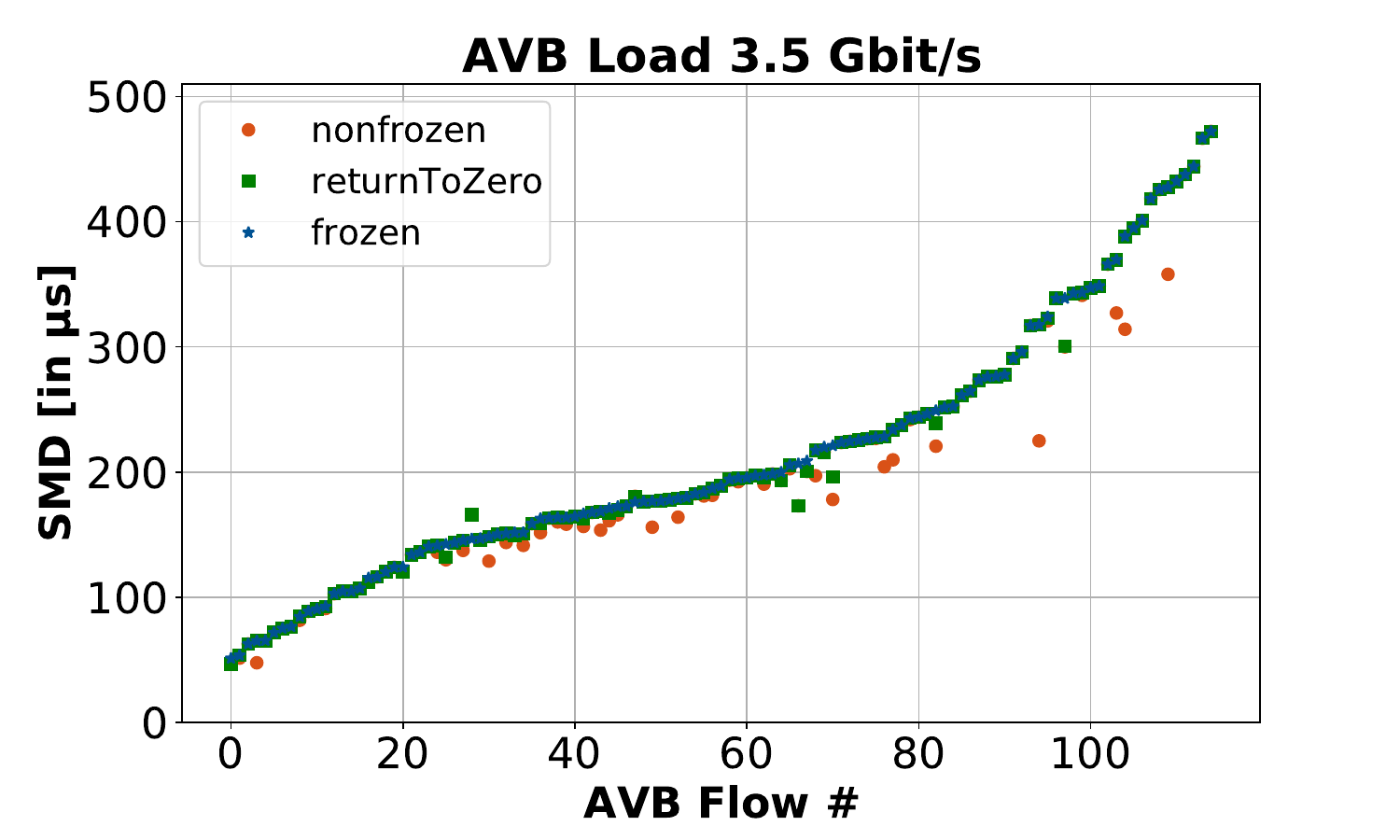}
  \caption{\textbf{ORION} Topology: \textbf{SMD} of AVB Flows, \textbf{CBS with GCL}, \textit{idleslope} = 750 Mbps.}
\label{fig:orion_SMD_AVBA_50_100_nonfrozen_frozen_returnToZero}
\vspace{-0.3cm}
\end{figure}

\begin{figure}[t!]
    \centering
    \includegraphics[scale=0.23, trim={0.5cm 0.5cm 3cm 1.2cm}, clip]{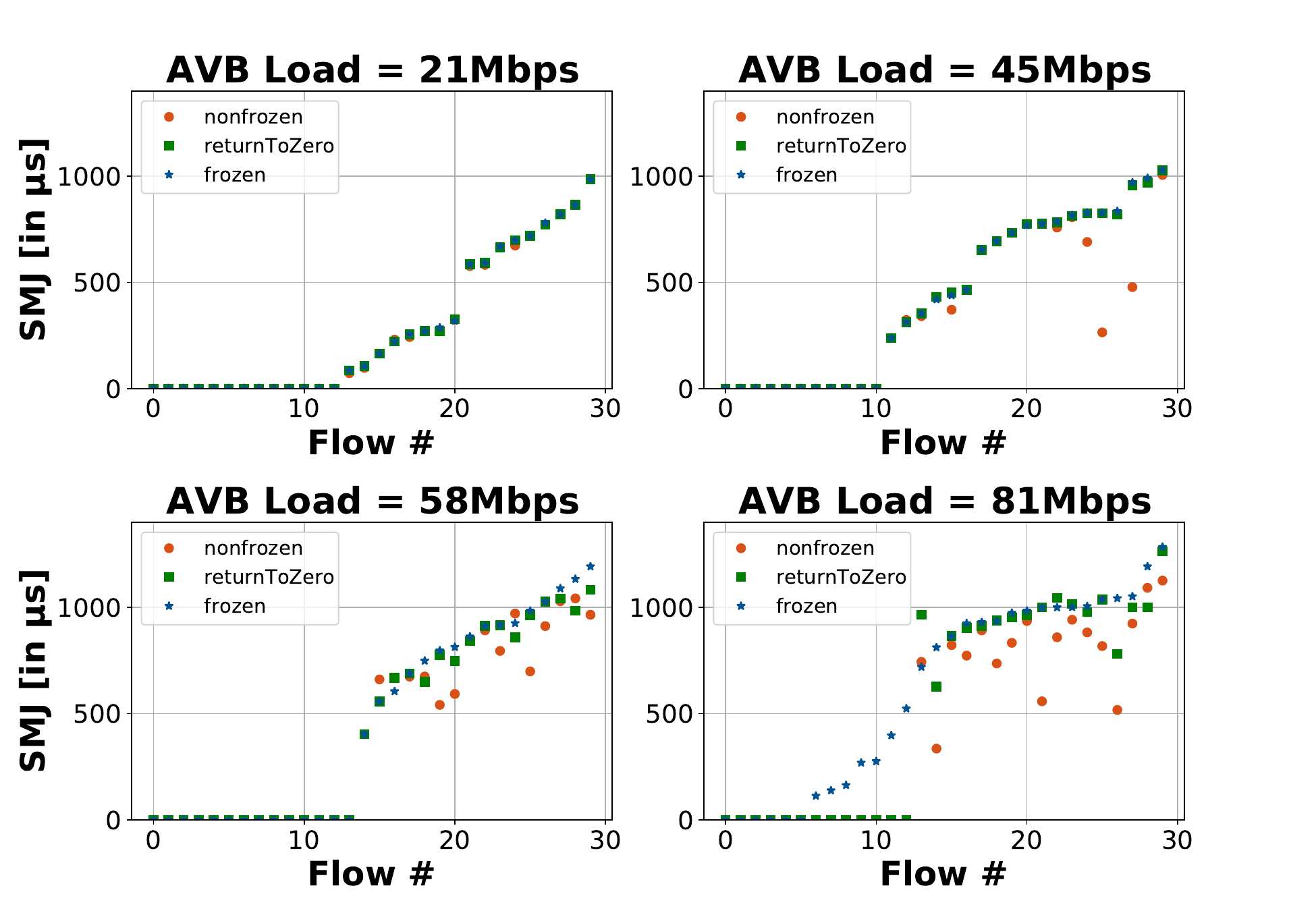}
    \caption{\textbf{MM}: \textbf{SMJ} of AVB Flows for \textbf{CBS with GCL.}}
\label{fig:max_all_max_SMJ_comparison_AVBA_30_40_50_60_frozen_nonfrozen}
    \vspace{-0.55cm}
\end{figure}

\subsection{Assessment of CBS with GCL}
\subsubsection{SMD}
Fig.~\ref{fig:SMD_AVBA_30to60_frozen_nonfrozen} displays the SMD values of the AVB traffic under three different integration modes for various network loads. As documented in previous research like \cite{boyer-cbs-frozen}, the "frozen" credit behavior generates the highest SMD values. A comparison of the three integration modes under different loads reveals that "nonfrozen" and "return-to-zero" generate lower SMD values compared to "frozen". It is evident from the SMD values that the "nonfrozen" credit mode is less affected at higher network loads. Similar behavior is observed in the industrial test case using the Orion topology, as shown in Fig.~\ref{fig:orion_SMD_AVBA_50_100_nonfrozen_frozen_returnToZero}. However, the performance differences between the various credit integration modes are minimal in the Orion test case. This is because Orion utilizes 1 Gbps link speeds, and the AVB flows experience less interference from high-priority traffic types. Additionally, in the Orion topology, the probability of frames encountering the infrequent "GBs" when the AVB load is low is very small. \textbf{\textit{In conclusion, the superiority of the three different integration modes becomes apparent when frequent gate opening and closing events cause a higher number of GBs, especially when the AVB load is substantial.}}

\subsubsection{SMJ}
\begin{figure}[t!]
    \centering
    \includegraphics[scale=0.34, trim={1cm 0.7cm 2cm 1cm}, clip]{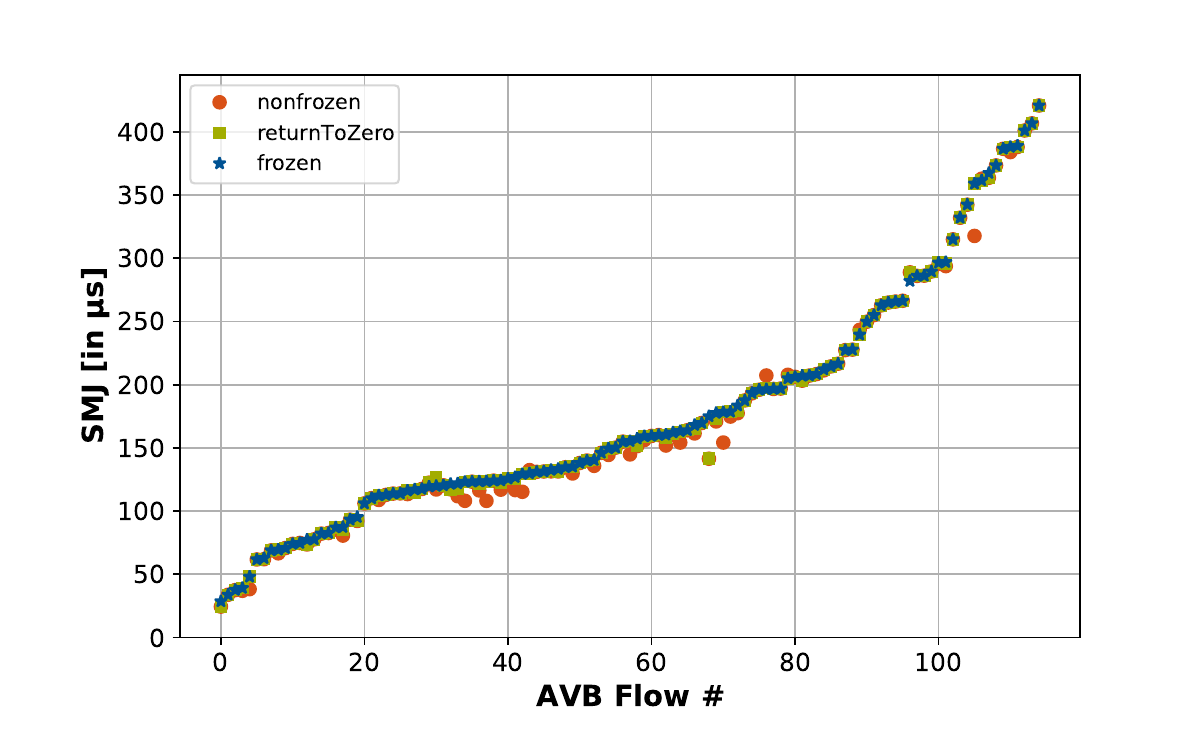}
    \caption{\textbf{ORION} Topology: \textbf{SMJ} of AVB Flows.}
    \label{fig:orion_SMJ_AVBA_nonfrozen_frozen_returnToZero}
    \vspace{-0.4cm}
\end{figure}

Fig.~\ref{fig:max_all_max_SMJ_comparison_AVBA_30_40_50_60_frozen_nonfrozen} illustrates the SMJ for all AVB load test cases in the MM topology. As the load increases, the SMJ also rises for the AVB flows. Notably, the SMJ is higher for the "frozen" and "return-to-zero" modes. Conversely, in the Orion test case, the SMJ values for all three integration modes are quite close to each other. Nevertheless, it is essential to note that none of the AVB flows achieve zero SMJ in the Orion test case, as shown in Fig.~\ref{fig:orion_SMJ_AVBA_nonfrozen_frozen_returnToZero}. However, much like in the MM test case, the "nonfrozen" mode exhibits the least variation in jitter. \textbf{\textit{In summary, the "nonfrozen" mode results in the least jitter variation.}}

 \begin{figure}[t!]
    \centering
    \includegraphics[scale=0.25, trim={0cm 0.4cm 0cm 0cm}, clip]{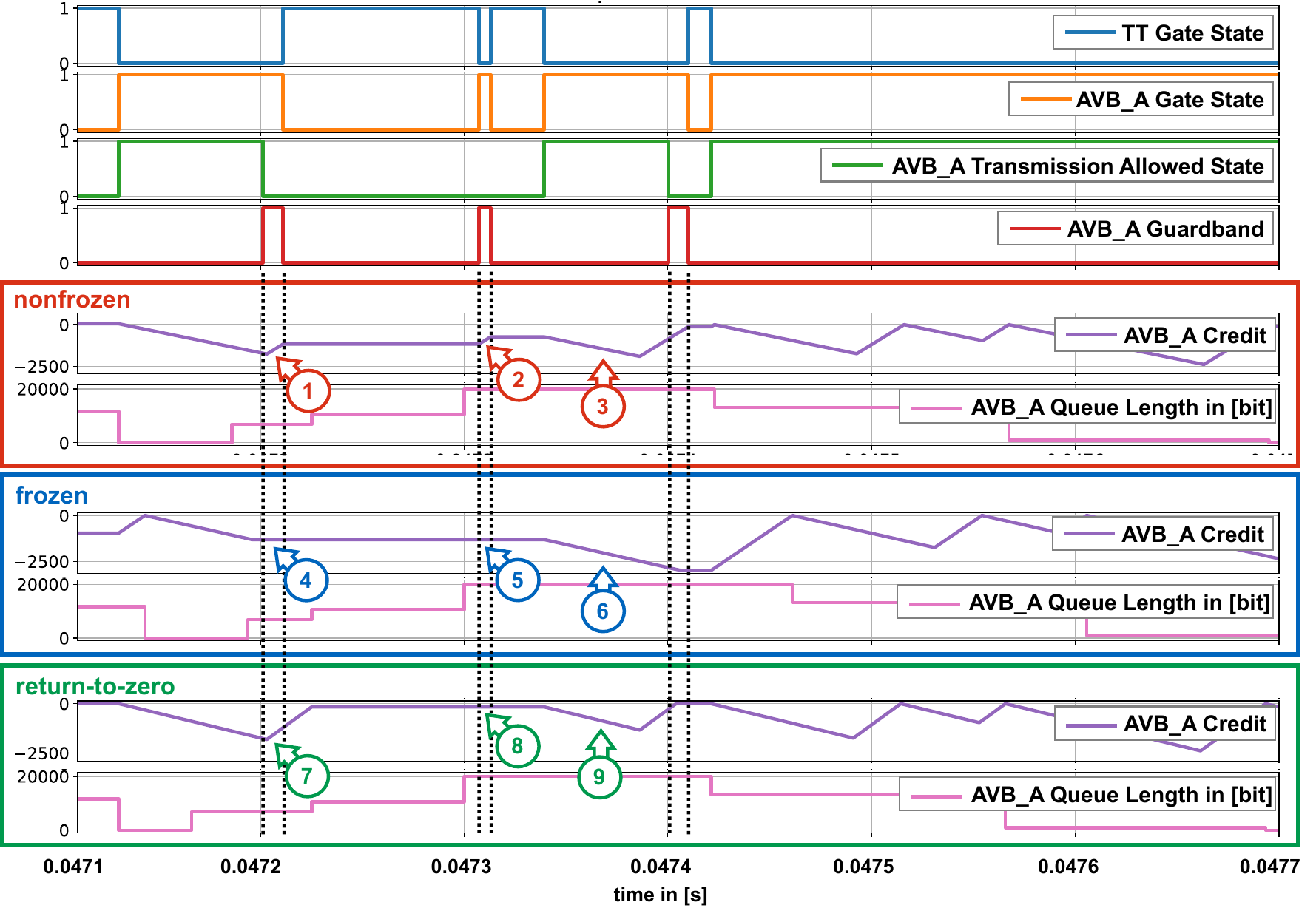}
    \caption{Credit Evolution FP with CBS with GCL \textbf{with} Hold/Release.}
    \label{fig:credit_evolution_TAS_CBS_FP_withHR}
    \vspace{-0.6cm}
\end{figure}

\begin{figure}[t!]
    \centering
    \includegraphics[scale=0.25, trim={0.3cm 0.5cm 2.5cm 0.5cm}, clip]{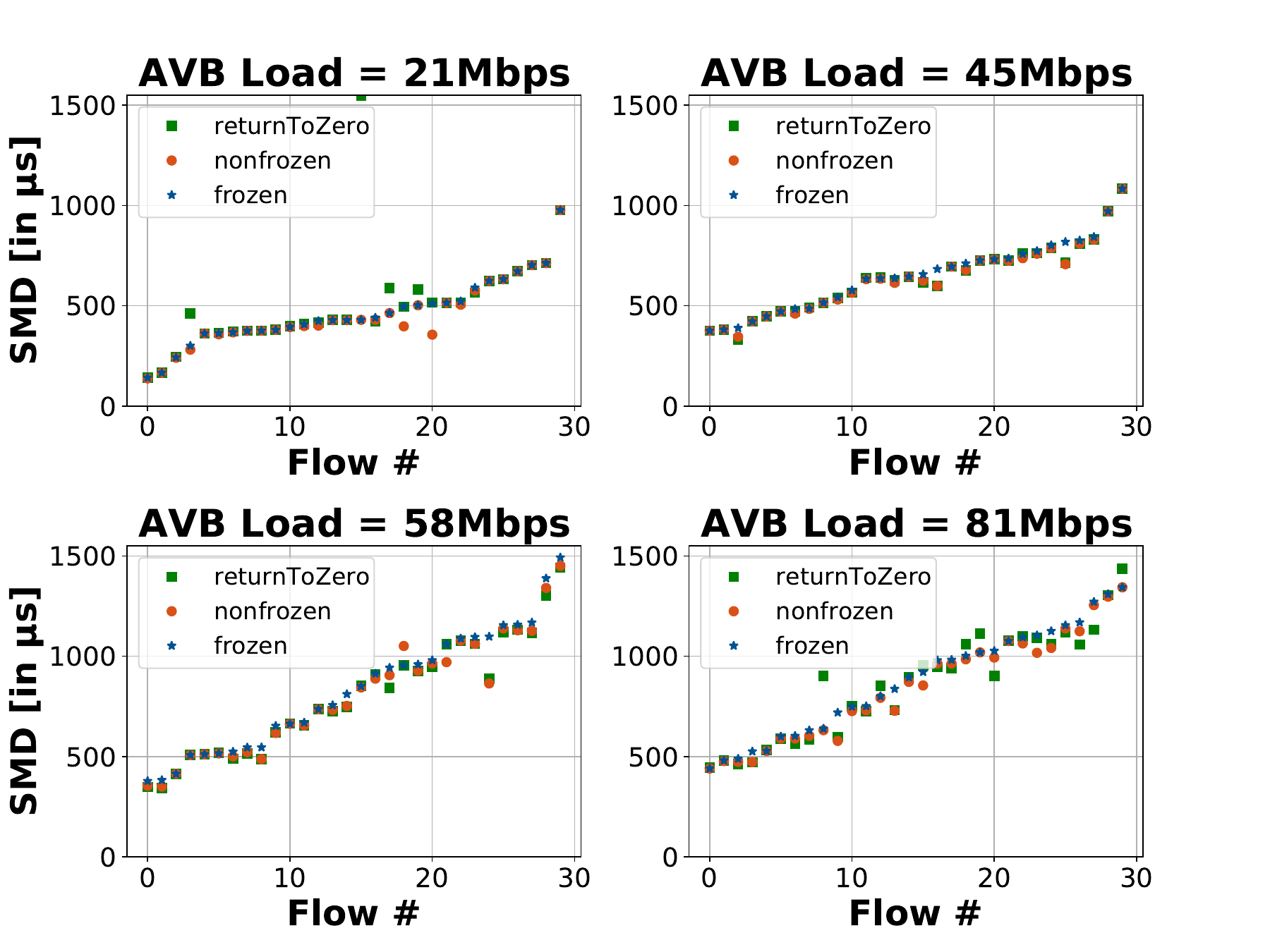}
    \caption{\textbf{MM} Topology: \textbf{SMD} of AVB Flows for \textbf{FP with CBS with GCL} \textbf{with} Hold/Release.}
    \label{fig:max_SMD_TAS_CBS_FP_ST_AVBA_frozen_nonfrozen_returnToZero}
    \vspace{-0.6cm}
\end{figure}

\subsection{Integration of FP with CBS and GCL}
In this sub-section, we assess the effects of employing FP for AVB flows that are shaped by CBS in a network featuring TT traffic and GCL. 

\begin{figure*}[t!]
    \centering
    \includegraphics[scale=0.35, trim={1.4cm 10.5cm 3cm 1.3cm}, clip]{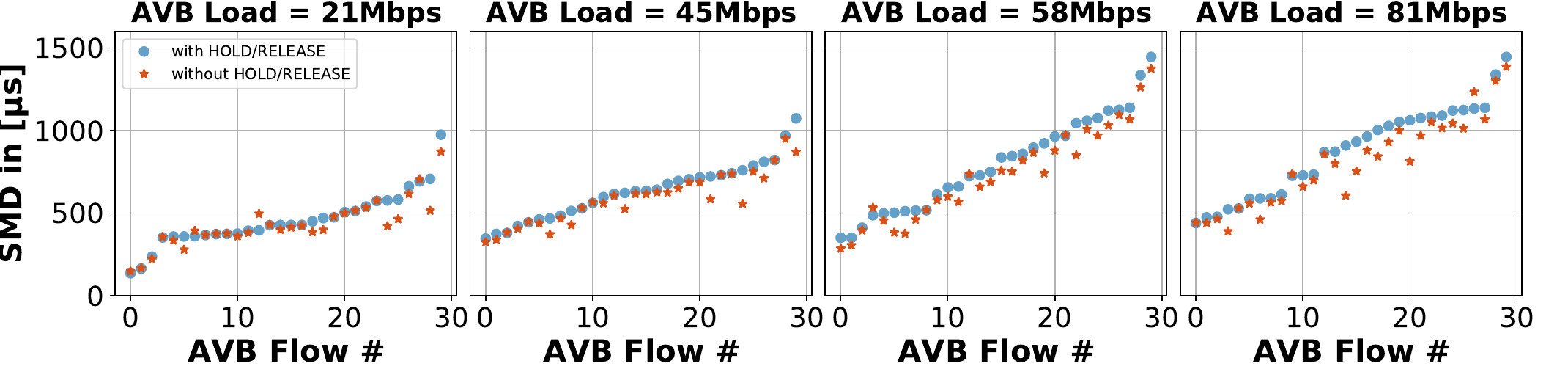}
    \caption{\textbf{MM} Topology: \textbf{SMD Comparison} of 30 AVB Flows for FP with CBS with GCL \textbf{with} and \textbf{without} Hold/Release.}
    \label{fig:max_SMD_comparison_TAS_CBS_FP_ST_AVB_A_standard_withoutHR_withHR}
    \vspace{-0.55cm}
\end{figure*}

\begin{figure}[t!]
    \centering
    \includegraphics[scale=0.32,trim={1cm 0.3cm 18cm 1.2cm}, clip]{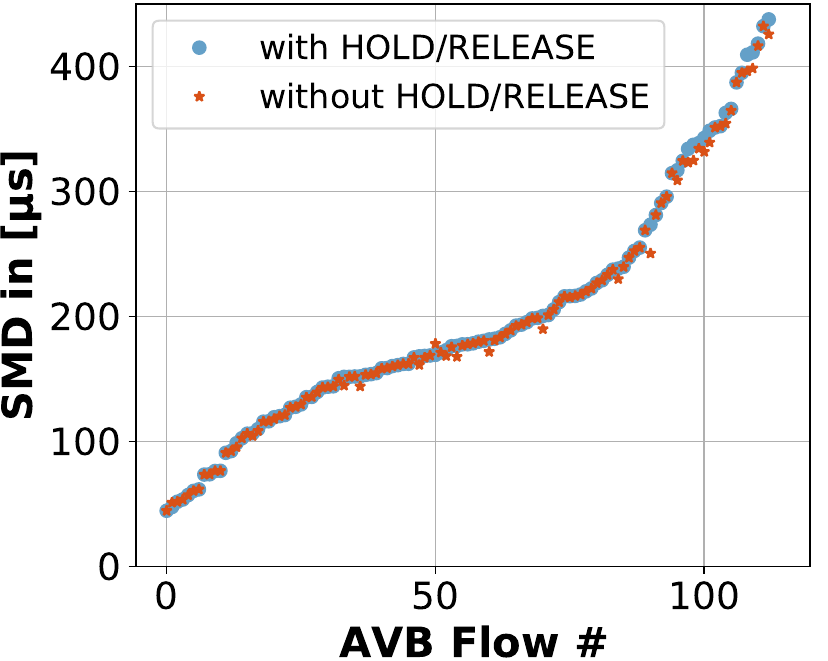}
    \caption{\textbf{ORION} Topology: \textbf{SMD} of AVB Flows for FP with CBS with GCL in \textbf{with} and \textbf{without} Hold/Release.}
    \label{fig:orion_max_SMD_comparison_TAS_CBS_FP_ST_AVB_A_standard_withoutHR_withHR}
    \vspace{-0.6cm}
\end{figure}

\subsubsection{FP with CBS with GCL in "with Hold/Release"}
Fig.~\ref{fig:credit_evolution_TAS_CBS_FP_withHR} illustrates the "credit" evolution curve of FP with CBS and GCL "with Hold/Release" for the "frozen," "nonfrozen," and "return-to-zero" modes. In the "nonfrozen" mode, indicated by the red box, FP with Hold/Release is initiated when the GB becomes active, triggering the "Hold" mechanism. This mechanism occurs before the AVB gate closing event, marked as "1". During this time, the credit value increases with the "idleSlope." When the AVB gate closes, credit accumulation pauses. Upon reopening the AVB gates, the GB immediately becomes active. This happens because the time between subsequent TT windows is less than 123 Bytes + Header length, causing a fragment to be unable to transmit and hence needs to wait. The credit continues to increase during this GB period ("nonfrozen"), marked as "2". In the next gate opening duration, the previously preempted fragment (waiting since "1") is fully transmitted, marked as "3".

\subsection{Impact of "FP with Hold/Release" on AVB Traffic}
\label{sec:TAS_CBS_FP_HR_comparison}
Fig.~\ref{fig:max_SMD_TAS_CBS_FP_ST_AVBA_frozen_nonfrozen_returnToZero} shows the SMD of AVB flows while using "FP with Hold/Release". The SMDs of AVB flows are significantly lower for all credit integration modes than the SMD of AVB traffic shown in Fig.~\ref{fig:SMD_AVBA_30to60_frozen_nonfrozen}. In Fig.~\ref{fig:SMD_AVBA_30to60_frozen_nonfrozen}, maximum SMD value of AVB flow was $2500\mu s$. However, when AVB flows are shaped using CBS but set to preemptable class (use of FP with CBS), the SMD value went down to $1500\mu s$ for the highest network load. Furthermore, the difference of SMDs between different GCL and CBS credit integration modes decreased significantly. Here the influence of the shortened GB to 123 Byte becomes prominent.

\subsection{Comparison between FP "With Hold/Release" vs. FP "Without Hold/Release"}
\label{sec:With_without_comparison}
Fig.~\ref{fig:max_SMD_comparison_TAS_CBS_FP_ST_AVB_A_standard_withoutHR_withHR} shows the comparison between the SMD values of AVB flows for two FP integration modes (\textit{with and without Hold/Release}) with CBS and GCL in the network for different AVB loads. The SMD values of AVB flows are smaller for "without Hold/Release" compared to "with Hold/Release". For the Orion network, the comparison between the with and without Hold/Release mode is shown in Fig.~\ref{fig:orion_max_SMD_comparison_TAS_CBS_FP_ST_AVB_A_standard_withoutHR_withHR}. Although, the difference between the SMD values for both of the FP integration modes is not as significant as the MM topology, we can still see the positive difference between the two modes for the AVB flows concluding that the SMD of the AVB flows is small for "without Hold/Release". \textbf{Before concluding, it is crucial to perform a performance evaluation of the impact of FP "with Hold/Release" and "without Hold/Release" on TT flows.}

\vspace{-0.5mm}
\subsection{Impact of with and without Hold/Release on TT traffic}
\begin{figure*}[t!]
    \centering
    \includegraphics[scale=0.35, trim={1cm 10.5cm 2.5cm 1cm}, clip]{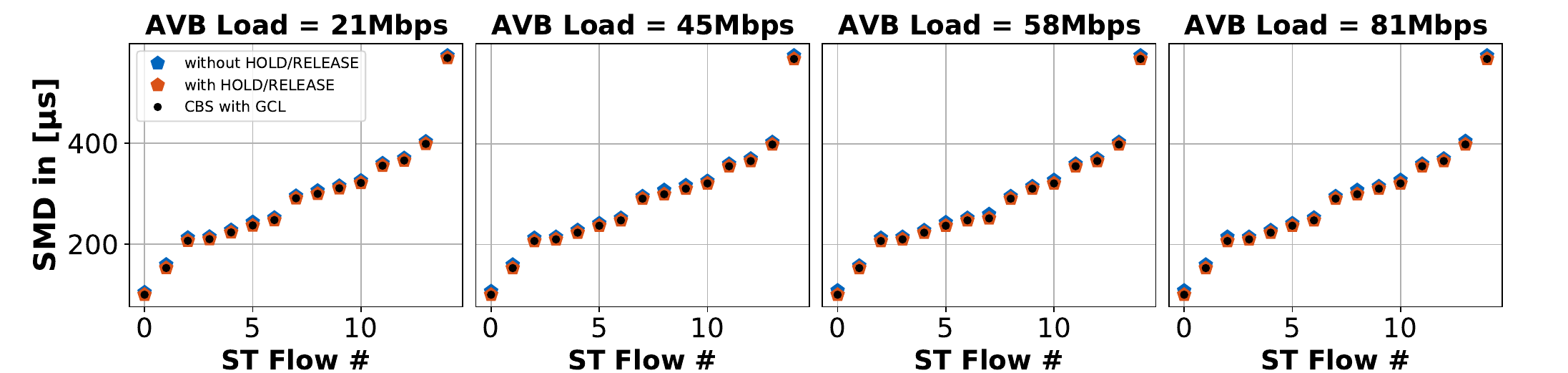}
    \caption{\textbf{MM} Topology: \textbf{SMD} of TT Flows; FP with CBS with GCL in \textbf{with} and \textbf{without} Hold/Release vs. CBS with GCL.}
    \label{fig:tt_delay_plot_all_comparison}
    \vspace{-0.55cm}
\end{figure*}

\begin{figure*}[t!]
    \centering
    \includegraphics[scale=0.35, trim={2cm 10.5cm 2.5cm 1cm}, clip]{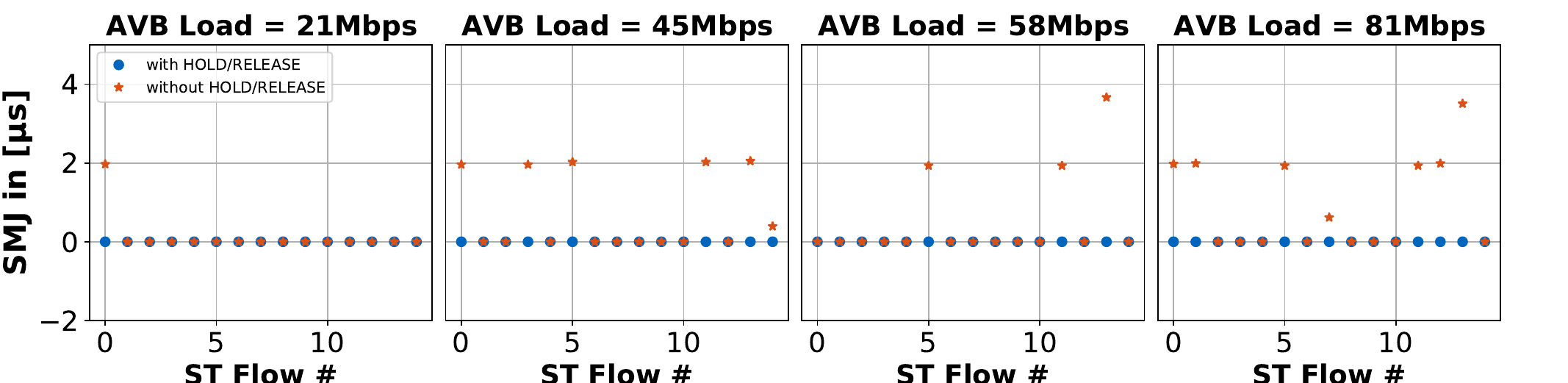}
    \caption{\textbf{MM} Topology: \textbf{SMJ Comparison} of 15 TT Flows; FP with CBS with GCL in \textbf{with} and \textbf{without} Hold/Release.}
    \label{fig:tt_jitter_plot}
    \vspace{-0.55cm}
\end{figure*}

Fig.~\ref{fig:tt_delay_plot_all_comparison} and Fig.~\ref{fig:tt_jitter_plot} shows the delay and the jitter of the TT flows for the with and without Hold/Release integration mode respectively. From the results, between the two integration modes, the "with Hold/Release" mode provides zero jitter for TT traffic. However, the jitter of the TT flows is high when we select the "without Hold/Release" mode. Furthermore, the SMD of TT flows are the same for the CBS with GCL vs. the FP with CBS with GCL in "with Hold/Release". The SMD of TT flow increases in the FP with CBS with GCL in "without Hold/Release" mode. As the proposed "without Hold/Release" in our paper includes a GB, which results in relatively low TT flow jitter. It is important to note that without the proposed GB in the "without Hold/Release" mode, both the SMJ and the SMD of TT flows would increase further. \textbf{In summary, "with Hold/Release" mode ensures zero TT flow jitter at the expense of increased SMD for AVB flows, while "without Hold/Release" introduces higher SMD and SMJ for TT flows while reducing the SMD of AVB flows.}

\vspace{-0.5mm}
\subsection{FP with CBS with GCL vs. CBS with GCL}
\begin{figure*}[t!]
    \centering
    \includegraphics[scale=0.35, trim={1cm 10.5cm 3cm 1cm}, clip]{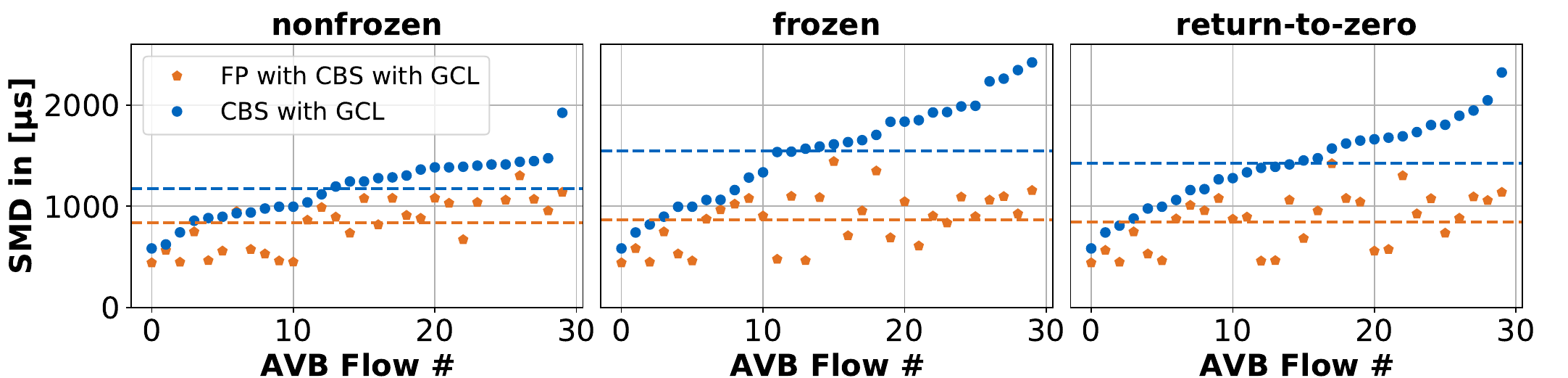}
    \caption{\textbf{MM} Topology: \textbf{SMD} of AVB Flows for \textbf{CBS with GCL} vs. \textbf{FP with CBS with GCL with Hold/Release}.}
    \label{fig:max_SMD_comparison_TAS_CBS_with_withoutFP_ST_AVB_A_60_standard_frozen_returnToZero}
    \vspace{-0.5cm}
\end{figure*}

\begin{figure*}[h!]
    \centering
    \includegraphics[scale=0.35, trim={1cm 10.5cm 2cm 1cm}, clip]{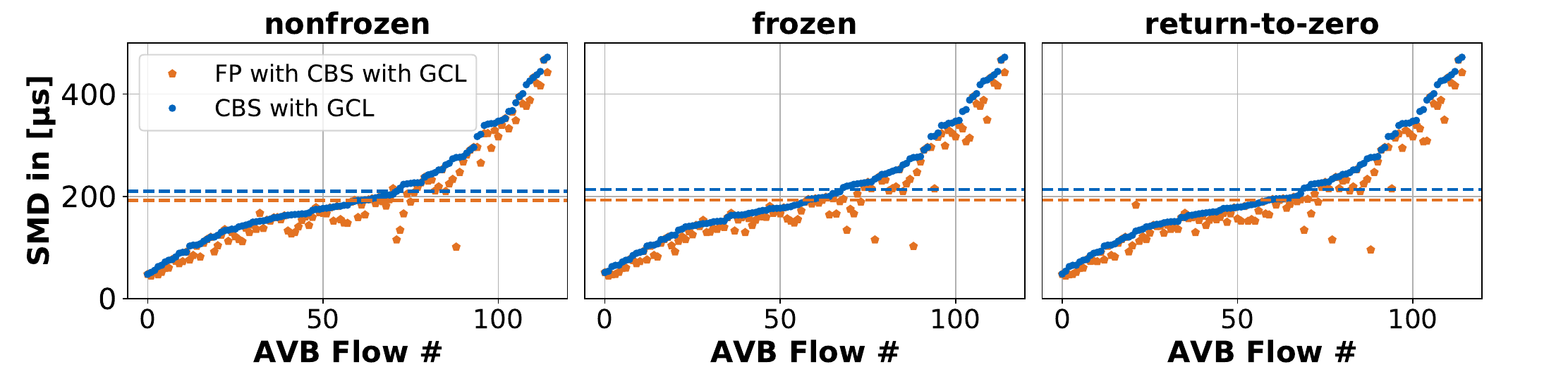}
    \caption{\textbf{Orion} Topology: \textbf{SMD} of AVB Flows for \textbf{CBS with GCL} vs. \textbf{FP with CBS with GCL with Hold/Release}.}
    \label{fig:orion_max_SMD_comparison_TAS_CBS_FP}
    \vspace{-0.55cm}
\end{figure*}

In our final comparison, we evaluate the SMDs of AVB flows in FP with CBS with GCL versus CBS with GCL. This evaluation unequivocally demonstrates the enhancement in AVB flow performance when preemption is employed alongside GCL. Importantly, we use the "with Hold/Release" mode to ensure TT flow performance remains unaffected. Fig.~\ref{fig:max_SMD_comparison_TAS_CBS_with_withoutFP_ST_AVB_A_60_standard_frozen_returnToZero} illustrates the SMD of AVB flows for the MM topology, and Fig.~\ref{fig:orion_max_SMD_comparison_TAS_CBS_FP} presents the contrast in SMDs for the Orion topology. \textbf{It is evident that the use of FP improves the performance of AVB flows and furthermore "with Hold/Release" has no impact of the TT flow performance.}

\section{Related Work}
\label{sec:related}
Arestova et al. presented the performance analysis of TAS and FP in \cite{anna_erlangen}. Support of multi-level preemption and the novel FP algorithms are further proposed in \cite{multi_level_preemption, rtns_multi_fp_ojewale, saad_novel_fp}. On the contrary, our work considers the impact of FP on the AVB flows under different integration modes together with GCL and further highlights the behaviour of the credit evolution when FP is used together with CBS. Ashjaei et al. in \cite{saad_novel_fp} proposed a novel FP mechanism which improves the worst-case response times of high priority frames compared to the state-of-the-art FP model. Debnath et al. in \cite{rubi} proposed the joint shaper architecture but did not consider the FP mechanism. Gavrilut et al. mentioned in \cite{voica} that the non-preemption integration mode leads to bandwidth wastage due to the GB. They further mentioned that the use of preemption will decrease the latency of AVB traffic and improve bandwidth usage; at the cost of slightly increased jitter for TT traffic. In our paper, we implemented and showed that AVB flows have lower delays when set to preemptable class, as the GB is reduced. Furthermore, the slightly increased jitter only happens when the FP integration mode is set to without Hold/Release. Ashjaei et al. in \cite{implications} showed different FP configuration and the timing behaviors of different traffic types mainly focusing on TAS, CBS and FP. However, our work is focused on the performance improvement of AVB when set as preemptable and the combined architecture support. Berisa et al. proposed the TT schedule synthesis with AVB and FP in \cite{rtns_fp_luxi}. They developed a Network Calculus solution for FP with Hold/Release with the goal to improve the schedulability of AVB flows. In our work, we used the TT traffic synthesis designed for CBS with GCL and showed the benefits of FP on AVB flows.

\vspace{-0.4cm}
\section{Conclusion}
\label{sec:conclusion}
In this paper, we have demonstrated the impact of FP on AVB flows within a FP architecture combined with CBS and GCL. Configuring AVB flows shaped by CBS in the FP with CBS and GCL architecture substantially reduces the end-to-end delay of AVB flows. Interestingly, the simulation performance of this combination has not been discussed in prior research. We have placed specific emphasis on assessing the performance of AVB and TT flows under various configurations and have evaluated all possible integration and FP mechanisms. One intriguing finding of this study is that the jitter of TT flows is higher for the "without Hold/Release" mode, even though the performance of AVB is better in this mode. Therefore, for optimal TT performance, the preferred integration mode for FP is "with Hold/Release", despite a slight increase in AVB delays. In summary, the results of our study offer valuable insights for network designers when selecting the appropriate FP combination with CBS and GCL.

\bibliographystyle{IEEEtran}
\bibliography{reference}
\end{document}